\documentclass[a4paper,useAMS, fleqn,usenatbib]{mnras} 
\usepackage[T1]{fontenc}
\usepackage{ae,aecompl}
\usepackage{graphicx}
\usepackage{animate} 
\usepackage{float}
\usepackage{siunitx}
\usepackage[caption = false]{subfig}
\usepackage[usenames, dvipsnames]{color}
\usepackage{hhline}
\usepackage{tikz}
\usepackage{multirow}
\usepackage{graphicx}
\usepackage{amsmath}	
\usepackage[utf8]{inputenc}
\usepackage{amssymb}
\usepackage[normalem]{ulem}
\usepackage{capt-of}
\usepackage{todonotes}

\newcommand{\nexus}{\textsc{nexus+}}
\newcommand{\massUnit}[1]{$ h^{-1} \rm{ M_{\odot}}$}
\newcommand{\distUnit}[1]{$ h^{-1} \; \rm {Mpc}$}
\newcommand{\mmf}{\textsc{mmf}}

\newcommand{\revised}[1]{\textcolor{black}{ #1}}

\newcommand{\eo}[1]{$\mathbf{e_1}$}
\newcommand{\et}[1]{$\mathbf{e_2}$}
\newcommand{\eg}[1]{$\mathbf{e_3}$}
\newcommand{\lcdm}{$\Lambda$CDM~}
\title[Evolution of halo spin alignment]{The Cosmic Ballet III: \\ halo spin evolution in the cosmic web}
\author[Ganeshaiah Veena et al.]
{\parbox{\textwidth}{
    Punyakoti {Ganeshaiah Veena},$^{1,2,3}$\thanks{E-mail:punyakoti.gv@gmail.com}
    Marius Cautun,$^{4,5}$ 
    Rien van de Weygaert$^{1}$ \\
    Elmo Tempel,$^{2}$ 
    and 
    Carlos S. Frenk$^{4}$
\vspace{.1cm} }
\\ 
$^{1}$Kapteyn Astronomical Institute, University of Groningen,PO Box 800, 9747 AD, Groningen, The Netherlands\\
$^{2}$Tartu Observatory, University of Tartu, Observatooriumi 1, 61602 T$\tilde{o}$ravere, Estonia \\
$^{3}$Department of Theoretical Physics, Tata Institute of Fundamental Research, Mumbai, 400005, India\\
$^{4}$Department of Physics, Institute for Computational Cosmology, University of Durham, South Road, Durham, DH1 3LE, UK\\
$^{5}$Leiden Observatory, Leiden University, PO Box 9513, NL-2300 RA Leiden, the Netherlands 
}

\pubyear{2020}

\begin{document}
\label{firstpage}
\pagerange{\pageref{firstpage}--\pageref{lastpage}}
\maketitle
\begin{abstract}
    We explore the evolution of halo spins in the cosmic web using a very large sample of dark matter haloes in the \lcdm Planck-Millennium N-body simulation. We use the \nexus{} multiscale formalism to identify the hierarchy of filaments and sheets of the cosmic web at several redshifts. We find that at all times the magnitude of halo spins correlates with the web environment, being largest in filaments, and, for the first time, we show that it also correlates with filament thickness as well as the angle between spin-orientation and the spine of the host filament. For example, massive haloes in thick filaments  spin faster than their counterparts in thin filaments, while for low-mass haloes the reverse is true. 
    We also have studied the evolution of alignment between halo spin orientations and the preferential axes of filaments and sheets. The alignment varies with halo mass, with the spins of low-mass haloes being predominantly along the filament spine, while those of high-mass haloes being predominantly perpendicular to the filament spine. On average, for all halo masses, halo spins become more perpendicular to the filament spine at later times. At all redshifts, the spin alignment shows a considerable variation with filament thickness, with the halo mass corresponding to the transition from parallel to perpendicular alignment varying by more than one order of magnitude. The \revised{cosmic web} environmental dependence of halo spin magnitude shows little evolution for $z\leq2$ and is likely a consequence of the correlations in the initial conditions or high redshift effects.
\end{abstract}

\section{Introduction}
Understanding the effects of the large-scale cosmic web on small-scale phenomena such as the growth of haloes and galaxies still remains an important open question in cosmology and galaxy formation. Besides small-scale processes such as AGN and supernovae, which have the largest impact on galaxy evolution, there is increasing evidence that processes on larger scales also play a role
\citep[e.g.][]{Dressler1980,Ball2008,Lewis2002,van_de_Weygaert2011,Beygu2016,Pandey2017}. Although the imprint of large scale on galaxy growth can be subtle, it needs to be studied such that we obtain a comprehensive understanding of
galaxy formation and cosmology. One of the prominent manifestations is the spin acquisition of haloes and galaxies and its connection to the cosmic web, which is yet to be completely understood. 
This represents the subject of this work.

According to the classical Tidal Torque Theory (TTT), angular momentum growth of a proto-halo is due to the large-scale tidal field. When the moment of inertia tensor of a proto-halo is misaligned with the surrounding tidal field, it experiences a torque and hence starts spinning. This was was first suggested by \cite{hoyle1949} and later studied in detail by \cite{peebles1969, doroshkevich1970, white1984}. The same tidal field is responsible for the anisotropic gravitational collapse of density fluctuations \citep{Peebles1980, Zeldovich1970, icke1973, vdw2008, desjacques2008} that result in the large-scale structure of the Universe, known as the cosmic web \citep[e.g.][]{bond1996,vdw1996, vdw2008}. 
The web represents the complex and hierarchical pattern seen in the large-scale distribution of matter, haloes, and galaxies, and consists of an intricate cellular structure composed of clusters, filaments, sheets and voids. 
The hierarchical nature of structure formation leads to 
numerous correlations between the spins of dark matter (DM) haloes and the cosmic web the haloes reside in. 
If the moment of inertia of a proto-halo and the surrounding tidal field are independent, then TTT predicts that the angular momentum of a halo is on average largest along the axis of second collapse \citep{lee2001, joneswey2009}, which is perpendicular to the filament spine and within the plane of the wall in which the halo is embedded. 
However, within the standard cosmological model the moment of inertia of a protohalo and the surrounding tidal field are  in-fact correlated
\citep{lee2000,porciani2002, porciani2002TTT2} and this, in turn, affects the orientation of halo spins. 
\citet{porciani2002} have shown that when accounting for the correlation between the inertia tensor and the initial tidal field, TTT predicts roughly equal alignment of the halo spin with the second and third eigenvectors of the initial tidal field. 

One manifestation of the effect of tidal fields on halo and galaxy spins is the alignment of the spin with the orientations of the cosmic web component in which the galaxies and haloes reside.
This correlation has been detected in both cosmological simulations and observations and it is a subject of active research in recent times due to a surge of available data. 
For example, cosmological simulations have found that there is a mass dependent alignment trend between halo spin and filament axis, with low-mass haloes having a propensity for parallel alignment with the filament axis and massive haloes spinning preferentially perpendicular to the filament axis \citep[e.g.][]{aragon2007, aragonPhd2007, hahn2007,  trowland2013, codis2012, codis2015, Libeskind2012, aragon2014, Welker2014, romero2014, Wang2017a, Wang2018a, Codis2018, GVeena2018, GVeena2019}. This mass dependent alignment is well described by the \citet{Lee2019} parametric model.

The mass at which the halo spin alignment changes from preferentially parallel to preferentially perpendicular is known as the transition mass. This is usually defined as the halo mass at which the median cosine of the angle between the spin vector and the host filament axis is 0.5, which marks random alignment.
While most studies have reported this transition in spin alignments, the value of the transition mass can vary by more than an order of magnitude between different studies. 
This is because the transition mass depends on the nature of filaments, with the transition mass being higher in thicker filaments \citep[this has been explicitly shown in][]{GVeena2018}. The filamentary network can vary between different web finders and this will be manifest as a different transition mass for the spin--filament alignment \citep[e.g. see][]{GVeena2018,GVeena2019}. The same effect is responsible for the transition mass varying with the smoothing scale used to identify the cosmic web \citep[e.g.][]{codis2012, aragon2014, romero2014}. 

Similar to haloes, the galaxies also show a mass dependent alignment between their spins and their host filaments. This has been shown in hydrodynamical simulation \citep[e.g.][]{Dubois2014,Welker2014,Wang2018b,GVeena2019,Kraljic2020} and also in observations. The first robust observational evidence was provided by the \citet{tempel2013} and \citet{TempelLibeskind2013} who have shown that the spins of spiral galaxies are preferentially aligned with the filament axis while the minor axis of elliptical galaxies, which are typically higher mass, is preferentially perpendicular to the filaments axis \citep[see also][]{jones2010, Hirv2017}. The same trend, although at a lower statistical significance due to the smaller sample, is seen when inferring the spin from the stellar or gaseous velocity maps, such as those obtained using SAMI or MaNGA \citep{Krolewski2019, Welker2020, bluebird2020}

The present day alignment between halo and galaxy spin and their filaments is different from that predicted by TTT. For example, as we just discussed, the high-mass haloes have a propensity for perpendicular spin while TTT predicts a parallel alignment. 
\revised{This mass trend can be qualitatively explained by anisotropic TTT \citep{codis2015} that takes into account that present-day filament haloes formed in particular Lagrangian tidal field configurations. However, being a Lagrangian theory, it does not capture the late-time non-linear stages of spin growth \citep{porciani2002}. The spin--filament alignment changes with time, especially at low redshift, \citep[e.g.][]{codis2012, Wang2017a, Wang2018b,Lopez2020a}.} 
It indicates that the spin orientation is affected by non-linear processes and that one of the manifestations of these processes is reflected in the spin--filament alignment and how it depends on halo, galaxy, and filament properties. This represents one of the key questions in the field, and
multiple ideas have been put-forth to explain it, such as: 
major merger events, vorticity generation inside filaments, formation and eventual migration of the halo into filaments and sheets, anisotropic accretion, and the connectivity of filaments \citep{codis2012, libeskind2013, Wang2018a,  Welker2014,  Laigle2015, romero2014, GVeena2018, GVeena2019}. 
Besides being essential for understanding halo and galaxy formation, the spin--filament alignment can be used to test cosmology, such as constraining the neutrino mass \citep{Lee2020}.

In this work, we build upon the \citet{GVeena2018} results, which investigated the halo spin--filament connection at $z=0$, and study as a function redshift the properties of DM halo spins and how they relate to the web component in which the halo resides. The goal is to determine the signatures of the non-linear processes that affect the halo spin growth and how these processes vary with the properties of the cosmic web. We do so by addressing the following questions:
\begin{enumerate}
    \item Does the halo spin magnitude depend on the cosmic web environment in which the halo is located?
    \item How does the halo spin--cosmic web alignment vary with cosmic time?
    \item How does the spin--filament alignment vary with filament properties at different cosmic times? 
    \item Is the magnitude of the halo spin correlated to the spin--filament alignment angle?
\end{enumerate}
\begin{figure*}
    \includegraphics[width=0.85\textwidth]{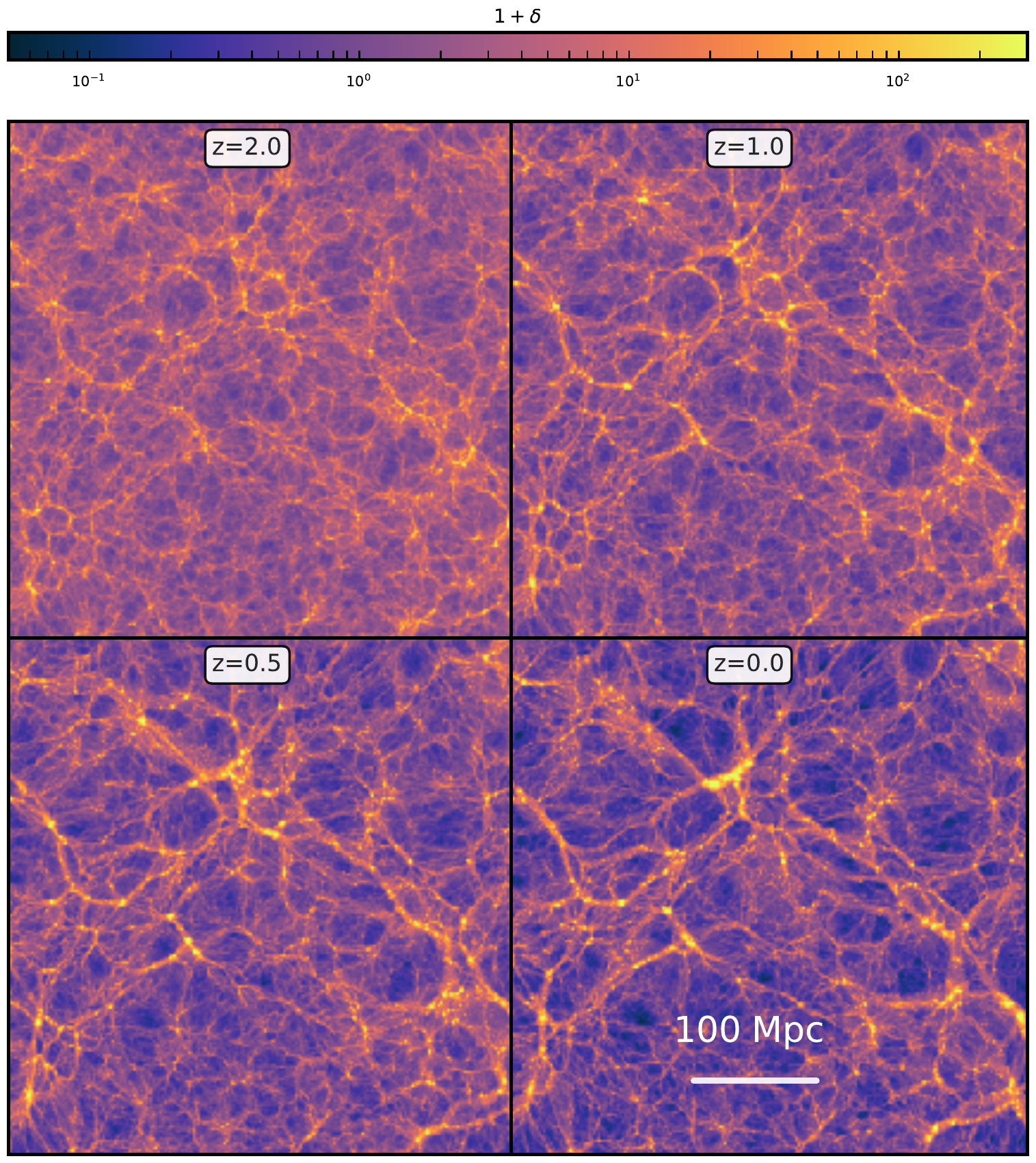}
    \caption{\textbf{Density field evolution:} The four panels show the density field of the P-Millennium simulation at different redshifts. Each plot is made using a slice of $2.3$\distUnit{}. The emergence of the cosmic web is clearly visible from these plots.
    }
    \label{fig:densityField}
\end{figure*}
To address these questions, we make use of a high resolution and large volume DM-only cosmological simulation, which allows us to identify the cosmic web in a representative region of the universe while also having resolved DM haloes over a wide range of masses. For each redshift of the simulation, we identify the population of haloes, defined as virialized collapsed regions, and the cosmic web. For the latter task, we use the \nexus{} method \citep{aragon2007MMF,cautun2013}; this is a multiscale approach that returns a hierarchy of filaments and sheets: from thick structures connecting the nodes of the web to tenuous ones in mostly underdense regions. Then, at each redshift we associate to a halo the web morphology and the web orientation identified at the halo's location. We then proceed to study correlations in the magnitude and direction of the DM halo spins with a halo's web morphology.

In our previous work, \citet{GVeena2018}, we have studied the halo spin--filament alignment at z=0 and its dependence on filament properties, such as filament thickness. In this paper, we study the evolution of the spin alignment of halo populations at different redshifts and explore to what extent this evolution varies for haloes residing in different filamentary environments, as quantified in terms of filament width.

The layout of the paper is as follows: \autoref{sec:PMill_simulation} contains the details of the simulation, halo population and selection criteria used for the study; in \autoref{sec:analysis} we describe how the spin alignment analysis is carried out; \autoref{sec:filamentAlignment} studies the evolution of the halo spin alignment with filaments and walls; in \autoref{sec:filamentThickness} we investigate the spin alignment in filaments of different thickness and how it varies with redshift; and finally, \autoref{sec:discussions} presents a short discussion and conclusions. 

\section{Filament and halo population}
\label{sec:PMill_simulation}
In order to address the question we mentioned above, we require an N-body simulation with a large number of well-resolved haloes and a large box size that is representative of the universe. For these reasons, we use the Planck-Millennium simulation of structure formation in a \lcdm cosmology. 

\subsection{P-Millennium simulation}
The Planck-Millennium  (or P-Millennium; \citealt{McCullagh2017,Baugh2019}) is a DM-only simulation of structure formation in a \lcdm cosmology.
It follows the evolution of 128 billion ($5040^3$) DM particles inside a 800 Mpc (542.16 \distUnit{}) box. 
The large box size combined with the high resolution makes it ideal to explore the evolution of halo properties in the cosmic web. The simulation employs the \citet{planck2014} cosmological parameters  and has a volume similar to the ground breaking Millennium simulation \citep{Springel2005a}, hence the name Planck-Millennium. The cosmological parameters used by the simulation are as follows: density parameters, $\Omega_{\Lambda} =0.693 $ and $ \Omega_{\rm M} =  0.307$, amplitude of the density fluctuations, $\sigma_8=0.8288$, and the Hubble parameter, $ h= 0.6777$, where $h = H_{0}/100$ $ \rm{ km \; s^{-1} Mpc^{-1}} $ and $H_0$ is the Hubble's constant at present day.

The simulation was run from $z= 127$ to present day, $z=0$. The initial conditions were generated using second order Lagrangian perturbation theory as described in \citet{Jenkins2010}. A total of 272 outputs or snapshots were generated, of which we have used four snapshots corresponding to $z= 2, 1, 0.5$ and 0.

\autoref{fig:densityField} is an illustration of the evolution of dark matter distribution from a redshift of 2 to 0 in the P-Millennium simulation. In this figure we plot the over-density given by, 
\begin{equation}
1+\delta({\mathbf{x}},t)\,=\,\frac{\rho({\mathbf{x}})}{\rho_u}
\end{equation}
where $\rho({\mathbf{x}})$ and $\rho_u$ denote the local and mean background density, respectively.
We plot this to show the formation and evolution of the cosmic web. At $z=2$ , the contrast of the structures is not yet very prominent, with some filaments and sheets clearly visible, but in general with a low contrast between high and low density regions. With time, the majority of web elements collapse and form highly dense nodes, elongated filaments and sheets, and large underdense volumes, i.e. voids. 
In each stage of evolution, matter flows from low to high density regions, which increases the density contrast. Filaments act as rivulets that transport matter from walls and voids into the high density cluster regions \citep{aragon2010, cautun2014, Buehlmann2019}. The smaller filaments at high redshift coalesce to form a more prominent filamentary network at later times. This is very neatly captured by the \nexus{} method (see Figure 21 in \citealt{cautun2014}).

\subsection{Halo population}
Haloes are found by first identifying Friends-of-Friends (FoF) groups \citep{Davis1985} using a linking length of 0.2 times the mean separation of dark matter particles (${\sim}0.16$ Mpc). Subsequently, the
gravitationally bound 
haloes 
are identified using the \textsc{subfind} algorithm \citep{springel2001}. It first detects the subhaloes associated to the local dark matter density peaks and then discards the particles that are not gravitationally bound to these substructures. \textsc{subfind} finds a hierarchy of subhaloes, with some being substructures of more massive subhaloes. For each FoF groups, the most massive object is defined as the main halo, and here we study only the main haloes. The halo centre is given as the DM particle with the lowest binding energy.

In this paper, we present the results using only the main \textsc{subfind} haloes and not the FoF groups. Since the FoF groups have multiple substructures linked together, the measurement of halo intrinsic spin may not be very meaningful in our context. In fact, we found that the spin distribution for FoF groups has a long tail of high values and that the spin distribution does not follow a log-normal distribution, especially at higher redshifts \revised{\cite{bett2007}}. 

We define the halo mass, $M_{200}$, as the mass of all DM particles enclosed within the radius $R_{200}$. The $R_{200}$ radius is that of a sphere centred at halo centre whose mean enclosed density is 200 times the critical density of the universe. For the halo spin, we use the values calculated by \textsc{subfind}, which consists of the angular momentum of all DM particles that are gravitationally bound to the halo. We use haloes with at least 300 DM particles, which are haloes with mass greater than $3.2 \times 10^{10}$\massUnit{}. With this criteria we have about $1.13 \times 10^{7}$ haloes at $z=0$ in the P-Millennium simulation. 

\begin{figure*}
    \setlength\tabcolsep{2pt}
    \begin{tabular}{@{} c c c @{}}
         \multicolumn{3}{c}{ \subfloat{\includegraphics[width=0.4\textwidth]{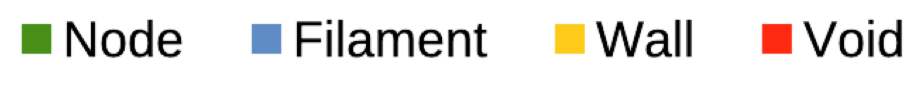}} } \\
         $ M_{200}=(3 - 5) \times 10^{10}$\massUnit{} &
         $M_{200}=(1 - 2) \times 10^{12}$ \massUnit{}&
         $M_{200}=(0.5 - 4) \times 10^{14}$\massUnit{} 
         \\[-.3cm]
         \subfloat{\includegraphics[width=0.32\textwidth]{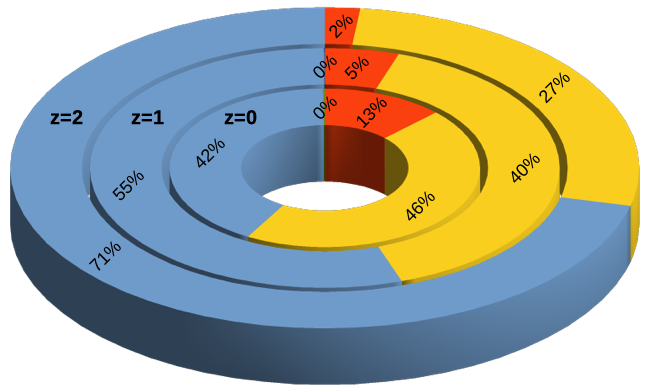}}
         &
        \subfloat{\includegraphics[width=0.32\textwidth]{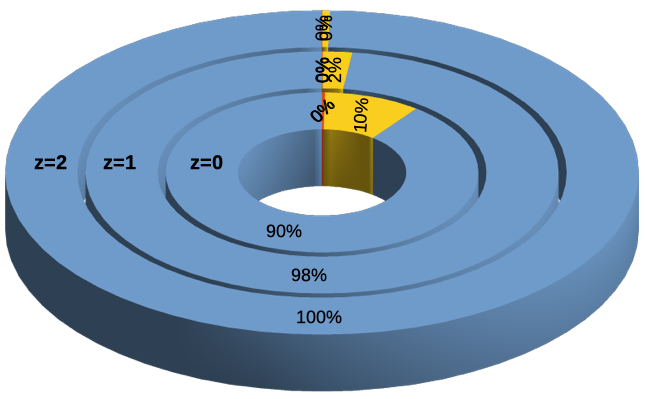}}
        &
        \subfloat{\includegraphics[width=0.32\textwidth]{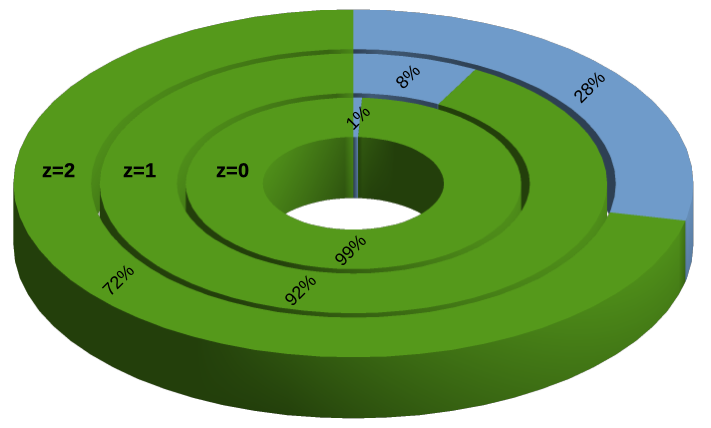}}
    \end{tabular}
    
    \caption{\textbf{Number fraction of haloes}: The panels show the fraction of main haloes in the different web environments. The outermost ring of the pie plot is for redshift $z=2$, the middle and inner rings are for $z=1$ and 0, respectively. The three panels show haloes of different masses: $(3.2 - 5) \times 10^{10}$\massUnit{} (left panel), $(1-2) \times 10^{12}$\massUnit{} (middle panel) and $(0.5-4) \times 10^{14}$\massUnit{} (right panel). The halo fraction in the various web environments varies rapidly with halo mass, with low-mass haloes residing mostly in filaments and walls, while high-mass ones are found mostly in nodes.}
    \label{fig:numberFrac}
    \end{figure*}
\begin{figure*}
    \subfloat{\includegraphics[width=\columnwidth, trim = {0.0cm 1.5cm 0.6cm 0}, clip] {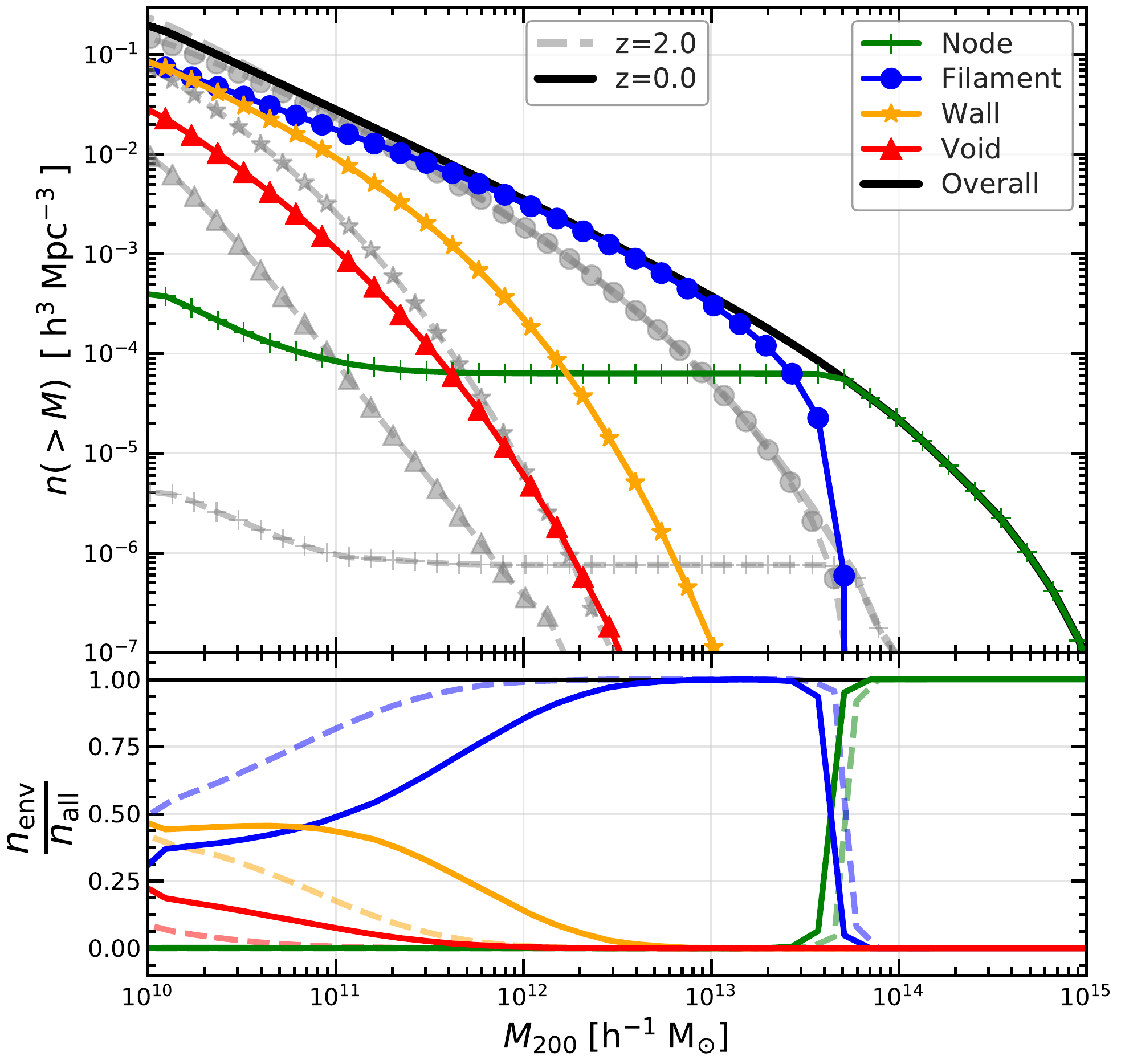}}
    \subfloat{\includegraphics[width =\columnwidth, trim = {0.0 1.5cm 0.6cm 0}, clip]
    {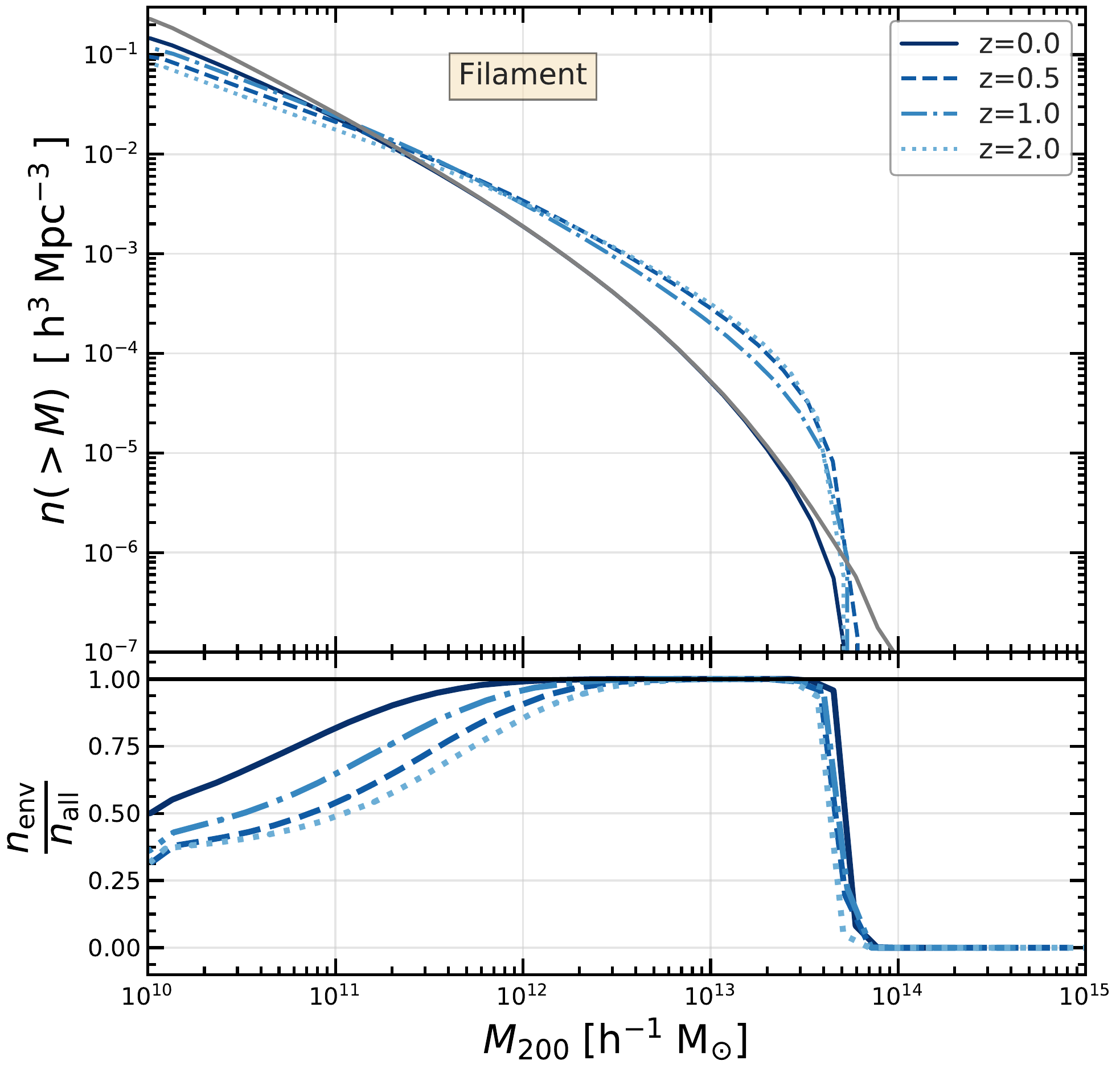}} \\
    \subfloat{\includegraphics[width =\columnwidth, trim = {0 0 0.6cm 0}, clip]{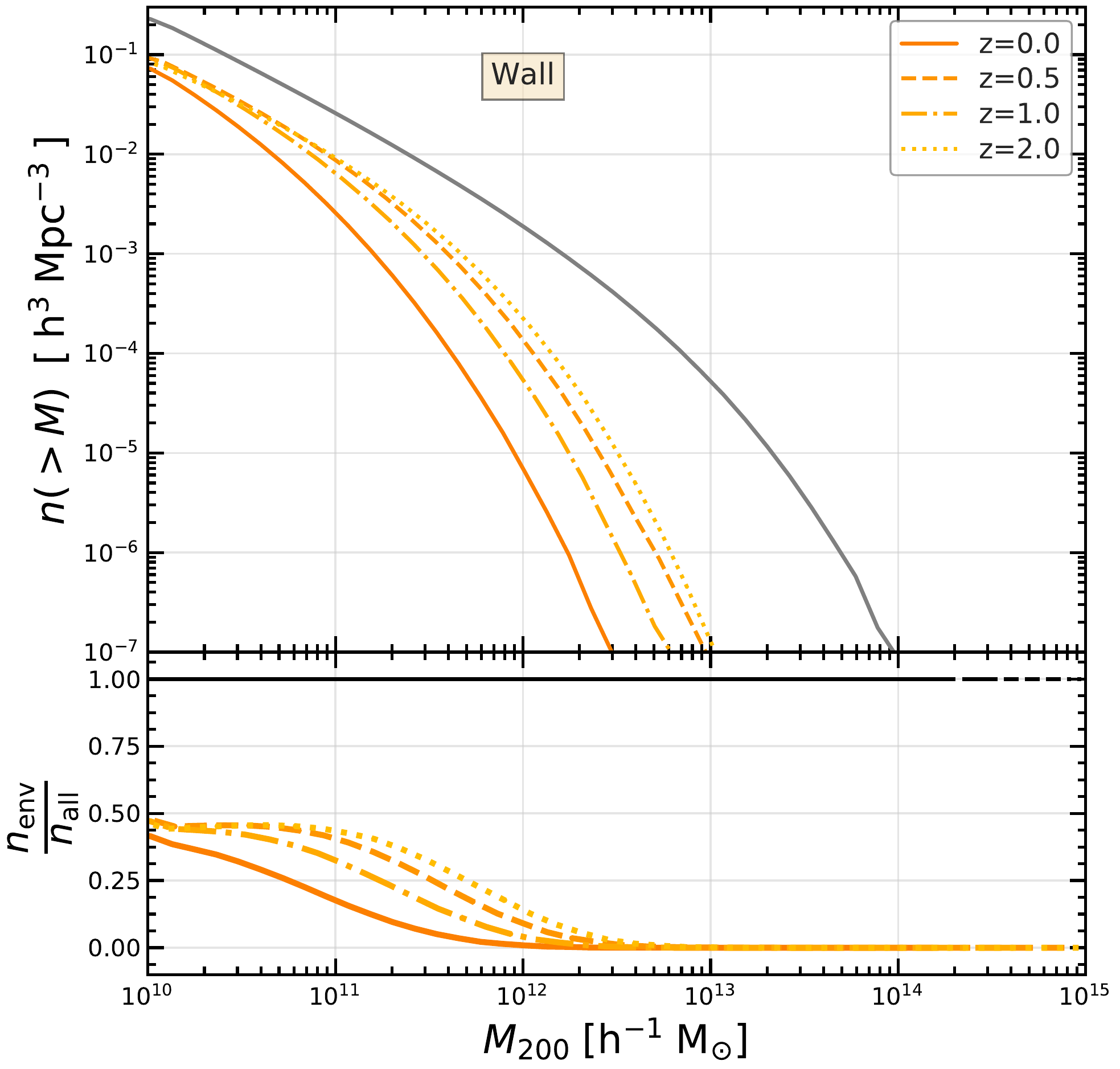}}
    \subfloat{\includegraphics[width =\columnwidth, trim = {0 0 0.6cm 0}, clip]{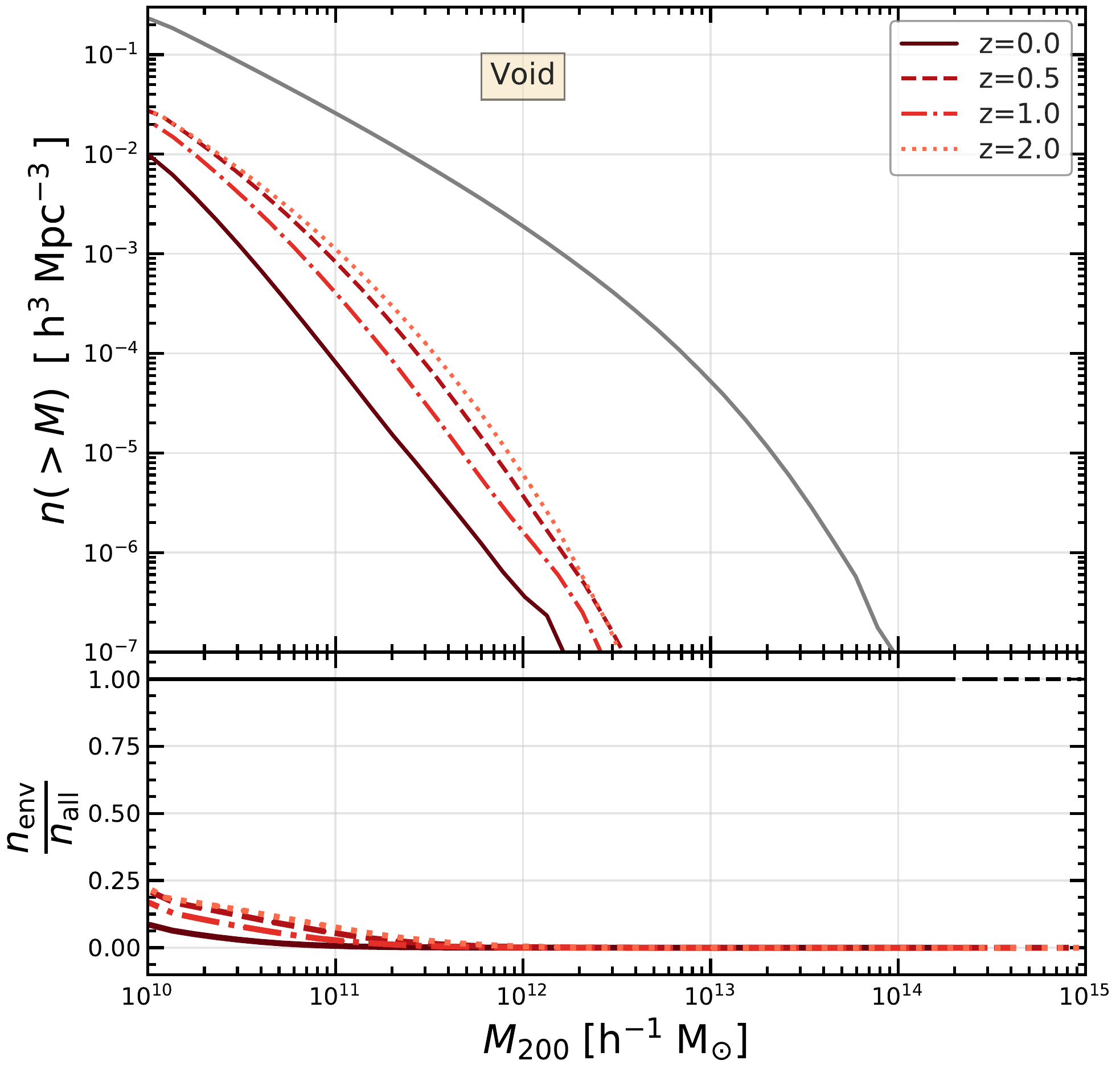}} 
    \caption{
    \textbf{Evolution of the halo mass function in the different web types:}
    \textit{The top-left panel} shows the cumulative halo mass function in the different web environments for redshifts, $z=0$ (coloured lines) and $z=2$ (light grey lines).
    The various colours are for haloes in different cosmic web environments at $z=0$. The differences of how haloes populate the various web environments are better captured in the bottom sub-panel, which shows the fraction of haloes in each web environment. \textit{The remaining panels} show the halo mass function in filaments (top-right), sheets (bottom-left) and voids (bottom-right) at four redshifts: $z=2, 1, 0.5$ and 0. For a sense of scale, the thin black line in each panel shows the halo mass function for the full halo population at $z=0$. Each panel has a bottom sub-panel that shows the fraction of haloes in that web type for the four redshifts.
    }
\label{fig:massFunctionEnv}
\end{figure*}

\subsection{Cosmic Web classification}
To characterise the morphological elements of the cosmic web in the simulation, we apply the \mmf{}/\nexus{} \citep{aragon2007MMF, cautun2013} method. The main feature of the method is that it identifies structures at all scales simultaneously based on the scale-space formalism. The multi-scale nature of the cosmic web is a consequence of the hierarchical structure formation and thus, to robustly identify all web elements, we need a multi-scale approach such as the one implemented within \nexus{}. 

The \nexus{} method takes as input a density field on a regular grid; for this we use a $1024^3$ grid (cell spacing of $0.53$\distUnit{}) and we calculate the density in each cell from the DM particle distribution using a nearest-grid-point assignment scheme.
\revised{Due to the very high number of DM particles of our simulation (on average ${\sim}5^3$ per cell), the majority of density cells contain one or more particles; for the very low fraction (${\sim}3\times10^{-5}$ of total at $z=0$) that do not contain any DM particles, we assign them a fiducial density corresponding to the cell containing half the mass of a DM particle. All these cells end up as part of voids and the exact density value we assign them does not impact the resulting cosmic web identification.}

The \nexus{} algorithm consists primarily of four steps or stages.
In the first stage, it calculates the logarithm of the input DM density field and 
smooths it with a Gaussian filter of different sizes. For implementing \nexus{} on P-Millennium, we have used a series of filter scales, where each scale is a factor of $\sqrt{2}$ larger than the previous one. The smallest scale we consider is 0.5 \distUnit{} (roughly the spacing of our grid cells), and the largest is 4.0 \distUnit{}, which is \revised{the largest smoothing scale that affects the filament identification within the \nexus{} framework} \citep{aragon2010, cautun2014}. \citet{cautun2014} have shown that going to smaller smoothing scales does affect the properties of filaments for $z\le 2$.
This is the crux of the scale-space approach where the data is represented at different filtering scales in order to capture the strongest features at each scale. 

In the second stage, the algorithm calculates the Hessian of the log-Gaussian filtered density field and obtains the eigenvalues of the Hessian matrix at every point. 
In the next stage, the eigenvalues ($\lambda_1 \leq  \lambda_2 \leq \lambda_3$) and eigenvectors ($\mathbf{e_1}, \mathbf{e_2} \;
\rm{and}\; \mathbf{e_3}$) are used to describe the local web morphology and orientation. The eigenvalues are used to calculate an environmental signature at each location. The exact calculation is a bit involved (see Eqs. 6 and 7 in \citealt{cautun2013}), but qualitatively \nexus{} defines the environments as follows. Filamentary structure is characterised by the condition that $\lambda_{1} \simeq \lambda_{2} < 0$ (matter is collapsing along two directions) and $ |\lambda_{2}| \gg |\lambda_{3}| $ (the change in density along the third direction is small compared to the change along the other two directions). \revised{The direction of the filament spine} is given by the eigenvector $\mathbf{e_3} $, as shown in \autoref{fig:schematicFila}.
 A wall or sheet on the other-hand is characterised by $ \lambda_{1} < 0$ (collapse along one direction) and  $|\lambda_{1}| \gg |\lambda_{2}| \simeq |\lambda_{3}|$ (the density hardly changes along the second and third directions). The eigenvector $\mathbf{e_{W1}} $ is the vector perpendicular to the plane of the wall as illustrated  in the lower panel of \autoref{fig:schematicFila}. 

A web environment of a given thickness shows the largest signature when filtering the density on the same scale as the width of the structure. This motivates the third step of \nexus{}, which consists of combining at each position the environmental signature of all smoothing scales and keeping only the largest value.

The final step of the \nexus{} method consists of identifying the regions that robustly can be characterized as being part of nodes, filaments, and walls. It consists of determining a threshold value for the environmental signature. For example, all regions with filament signatures larger than the threshold are identified as filaments.
For nodes, the threshold is determined by requiring that at least half of the nodes are virialised, that is their mean density is at least the virial value \citep[see][]{BryanVirialThreshold1998}.
For filaments and walls the detection threshold is decided automatically from the variation in the properties of the filamentary and wall network with environment signature (see appendix A in \citealt{cautun2013}). 
The regions that are not classified as nodes, filaments, or walls, are defined as voids.

Here we study the cosmic web from redshift, $z=2$, to present-day. To do so we analyze one at a time the snapshots of the P-Millennium simulation corresponding to $z=2, 1, 0.5$ and 0, i.e. we apply \nexus{} and cross-correlate with the halo catalogue separately for each snapshot. In principle, we may extend our study to even higher redshifts, however doing so comes with practical difficulties. The typical width of filaments and sheets decreases rapidly with redshift \citep{cautun2014} and to robustly trace them at higher redshifts we need to calculate the cosmic web using a finer grid. Currently, we use a $1024^3$ grid with a grid spacing of $0.53$\distUnit{}; further increasing the number of grid cells leads to a higher computational cost and especially RAM requirement.

\subsection{Evolution of halo mass function}
A simple way to quantify the effect of the cosmic web onto the halo population is to study how the halo mass function varies with web environment.
\autoref{fig:numberFrac} shows the evolution of number fraction of haloes in different web environments for three halo mass ranges. The fraction of low-mass haloes in filaments is high at $z=2$ and decreases towards present day, with the remaining haloes being mostly located in sheets and, a small fraction, in voids. A similar trend is observed for the intermediate mass haloes, $M_{200}\sim1\times10^{12}$\massUnit{}, but in this case only a small fraction ($<10\%$) is found in sheets and hardly any inside voids or nodes.  At even higher masses, most haloes are found in nodes and only a small fraction in filaments. 

The fraction of haloes in different web environments varies strongly with halo mass, as illustrated in \autoref{fig:numberFrac}. To have a comprehensive view of this dependence, in the top-right panel of \autoref{fig:massFunctionEnv} we show the cumulative number density of haloes, $n(>M_{200})$, as a function of halo mass, $M_{200}$, segregated by cosmic web type. To help with the interpretation of the plot, the bottom sub-panel shows the fraction of haloes in each web type. At present time, which is shown by the coloured lines, most haloes with $M_{200}>5 \times 10^{13}$\massUnit{} represent the nodes of the web, while most of the lower mass haloes are found in filaments. Sheet haloes become an appreciable fraction of the population for $M_{200}< 1 \times 10^{12}$\massUnit{} and void haloes become important at even lower masses, $M_{200}< 1 \times 10^{11}$\massUnit{} \citep{cautun2014}. Qualitatively, the picture is similar at $z=2$ (light grey lines), but with fewer haloes for a given mass especially in sheets and walls.

To better quantify the change in halo population with redshift, the remaining panels of \autoref{fig:massFunctionEnv} show the halo mass function in filaments, sheets and voids for the four redshifts we analyze in this paper. We typically find that at fixed halo mass we have more haloes in a given environment at $z=0$ than at $z=2$. This is the case for all web environments, except for filament haloes with $ M_{200}< 1 \times 10^{12}$\massUnit{}, in which case we observe the opposite trend. We also find that the halo mass function in sheets and voids increases the most towards present day, while for filaments the change is more modest. \revised{This makes sense, with haloes in filaments having formed earlier 
\citep[e.g.][]{hahn2007,Hellwing2020} than in sheets and voids, and thus evolving less at later times, than their equal mass counterparts in other web environments.}

In terms of halo fraction, we observe the following (see the bottom sub-panels for each panel in \autoref{fig:massFunctionEnv}). At the low-mass end, $ M_{200} \; \leq 1 \times 10^{12}$\massUnit{}, we find that at $z=2$, more than 50\% of the haloes are in filaments, followed by walls and voids regions. At the current time, this fraction is reduced in filaments and there is accordingly an increase in the wall and void fractions. The decrease in the number of low-mass filament haloes is because some objects merge to form more massive haloes at later times, hence we see a reduction in the low-mass end and increase in the high-mass end. 

\subsection{Characteristic halo mass}
Structure formation in \lcdm cosmologies proceeds hierarchically. Haloes build up by gradual merging of smaller haloes and the accretion of mass.
The first object to emerge are low-mass haloes, which subsequently grow into ever more massive structures. When comparing the halo populations at
different cosmic epochs, we need to take this process into account: an individual halo at a given redshift $z$ is the product of the growth of
a lower mass halo at higher redshift through merging and accretion. While in the present study we investigate the evolution of the entire
halo population, we incorporate the hierarchical growth of haloes by means of a characteristic halo mass at each
redshift.

A reasonable definition for the characteristic halo mass, $ M^{*}(z) $, follows from the analytical description of the hierarchical
process in terms of the Press-Schechter formalism \citep{PressSchecter1974,bondcole1991,laceycole1993}. It infers the expected number
density,  $n(M,z)$, of haloes of mass, $M$, at a redshift, $z$, assuming that structure emerges from an initial Gaussian
density field and that mass clumps on a mass scale $M$ would collapse if their linear extrapolated overdensity would surpass
the threshold level for gravitational collapse  \citep{Gunn1972}. In most practical circumstances, the collapse threshold is taken as $\delta_c=1.686$, which is the critical collapse overdensity of a spherical peak in an Einstein-de Sitter Universe
\citep[see eg.][for more realistic estimates]{shethtormen2001}.

The Press-Schechter mass function typically consists of a power-law low-mass wing that diverges to low masses, with an exponential
cut-off that reflects the Gaussian nature of the initial fluctuations. The characteristics halo mass, $M^{*}(z)$ , for this hierarchically evolving halo population is the exponential cut-off mass. It is the scale at which the field variance $ \sigma(M) $ on mass scale $M$ is equal to the critical linear over-density of gravitational collapse,
\begin{equation}
  \sigma(M^{*})\,\sim\, \delta_c \,\approx\, 1.686\,.
  \end{equation}
In a sense, it is the mass scale at which the average density peak in the primordial mass distribution undergoes gravitational collapse.

For a given (linearly extrpolated) power spectrum, $P(k,z)$, the mass variance $ \sigma(M,z) $ is
\begin{center}
    \begin{equation}
    \sigma^{2}(M,z) = \int_{0}^{\infty} \frac{dk}{2\pi^{2}} \; P(k,z) \; \tilde{W}_{TH}^{2}(\mathbf{k}R) \;  k^2 ,
    \end{equation}
\end{center}
where $\tilde{W}_{TH}(\mathbf{k}R)$ is the Fourier transform of the (tophat) window function of radius $R$. This is the radius of the
sphere that encloses the mass, $M = 4\pi/3 \; \rho_u(z) R^3$, where $\rho_u(z)$ is the mean density of the Universe at epoch $z$. 

\bigskip
The evolution of the characteristic mass is indicative of the hierarchical buildup of nonlinear structures. The increase of $M^*$ with time reflects the buildup and emergence of
more and more massive haloes in the evolving universe. We use the values of this characteristic ``collapse mass'', $M^*(z)$, as
a means to assess in how far observed trends in the various processes may be ascribed to the hierarchical growth of individual
halo masses with time and, hence, in how far additional processes may be involved. The typical values of $M^{*}(z)$ for 4 different
redshifts, for the \citet{planck2014} power spectrum parameters, are listed in \autoref{table:CharacterMass}.

 \begin{table}
     \centering
     \caption{\textbf{Characteristic halo mass, $M^{*}(z)$}. } 
     \label{table:CharacterMass}
     \renewcommand{\arraystretch}{1.2}
 	\begin{tabular}{ @{}c cc@{}} %
 		\hline
         \hline \\[-.35cm]
 		Redshift &  $M^{*}(z)$ &  \\
                &   [ \massUnit{} ]  & \\
         \hline \\[-.35cm]
 		0.0 & $ 4.3 \times 10^{13} $ & \\
 		0.5 & $ 1.4 \times 10^{12} $ & \\
 		1.0 & $ 6.0 \times 10^{12} $ & \\
 		2.0 & $ 5.9 \times 10^{11} $ &  \\
 		\hline
    \end{tabular}
    \renewcommand{\arraystretch}{1.0}
 \end{table}

\section{Analysis}
\label{sec:analysis}

\subsection{Halo angular momentum}
The angular momentum,  $\mathbf{J}$, of a halo with $N$ particles, is calculated as 
\begin{equation}
    \mathbf{J} =  \sum\limits_{k=1}^{N} m_k \left( \mathbf{r}_{k} \times \mathbf{v}_{k}  \right),
\end{equation}
where $m_k$, $\mathbf{r}_k$ and $\mathbf{v}_k$ are respectively the mass, position and velocity of the $k^{th}$ particle with respect to the centre of the halo. We sum all the DM particles that are gravitationally bound to the \textsc{subfind} main halo.

\subsubsection{Spin parameter} 
Physically, it is more useful the express the amplitude of the angular momentum, $J=|\mathbf{J}|$, in terms of the spin parameter, $\lambda$. The $\lambda$ parameter gives the degree of coherent rotation of any self-gravitating system, in our case a DM halo. A value closer to unity implies that the halo is mostly supported by rotation while a low value means that the halo is dispersion supported. 

The spin parameter was first introduced by \citet{peebles1969} and involves quantities, such as the energy of the system, that are rather involved to calculate. An alternative simpler version  was introduced by \citet[][see \citealt{mo_van_den_bosch_white_2010} for how this compares to Peebles' definition]{bullock2001} and it is given by
\begin{equation}
    \lambda = \frac{J}{\sqrt{2}M V_h R_h},
\end{equation}
where $V_h$ is the circular velocity at the halo radius, $R_h$.

At $z=0$ the spin parameter follows a log-normal distribution with a median value, $ \left< \lambda \right> \simeq 0.04 $, and it hardly varies with halo mass \citep{bett2007}; the same holds true also for the haloes in the P-Millennium simulation \citep{GVeena2018}.
This low value indicates that DM haloes are mostly dispersion supported rather than rotation supported. To compare, for disk dominated galaxies in hydrodynamical simulations that are supported by rotation, the median spin parameter is an order of magnitude higher \citep{GVeena2019}.

\subsection{Alignment analysis}
\label{sec:alignmentAnalysis}
We characterise the alignment between the halo angular momentum, $\mathbf{J}$, and its host environment orientation, $\mathbf{e_n}$, by the angle, $\theta_{\mathbf{J}, \mathbf{e_n}}$, between the two vectors. 
Physically, it is better to express the alignment in terms of the cosine of the alignment angle, i.e. $\cos \theta_{\mathbf{J}, \mathbf{e_n}}$, since in three dimensions the distribution of $\cos \theta$ between two randomly oriented vectors is uniform. We calculate the alignment as
\begin{equation}
    \centering
	 \cos \theta_{\mathbf{J}, \mathbf{e_n}} = \left| \frac{\mathbf{J} \cdot \mathbf{e_n}}{|\mathbf{J}| |\mathbf{e_n} |} \right|\,.
    \label{cos} 
\end{equation} 

We take the absolute value of the dot product since filaments have an orientation and both $\mathbf{e_3}$ and $\mathbf{-e_3}$ are valid and equivalent for our case. If a halo spin points along the direction of the cosmic web, then the cosine value is close to one, whereas if the halo spin is nearly perpendicular then the value of the cosine is close to zero. If there is no alignment, i.e. the two vectors are randomly oriented, then the distribution of $\cos \theta_{\mathbf{J ; e_n}}$  is uniform between 0 and 1. Any deviation from this expectation reflects the deviation from isotropic distribution or random alignment. 

\begin{figure}
\centering
\begin{tikzpicture}[scale=1.2]
\begin{scope}
\fill [gray,opacity=0.5] (0,-1.25) -- (-5.0,-1.25) arc (270:90:.5 and 1.25) -- (0,1.25) arc (90:270:-.5 and 1.25);
\draw[ultra thick] (0,0) ellipse (.5 and 1.25);
\draw[ultra thick] (0,-1.25) -- (-5.0,-1.25);
\draw[ultra thick] (-5.0,-1.25) arc (270:90:.5 and 1.25);
\draw [ultra thick, dashed] (-5.0,-1.25) arc (-90:90:0.5 and 1.25);
\draw[ultra thick] (-5.0,1.25) -- (0,1.25);
\end{scope}
\begin{scope}[xshift=-1 cm]
\draw[ultra thick, fill=blue!50!, rotate=-15, yshift=-0.35cm,xshift=0.2cm] (-1.5,0) ellipse (1.25 and 0.5);
\draw[thick, ->] (-1.5,0) to (-0.7,1.1);
\node[below] at (-0.5,0.9) {$\mathbf{J_{dm}}$};
\draw[ultra thick, ->] (-1.5,0) to (-.4,0);
\node[below] at (-0.4,-.1) {$\mathbf{{e}_3}$};
\draw[ultra thick, ->] (-1.5,0) to (-1.5,0.9);
\draw[ultra thick, ->] (-1.5,0) to (-2,-0.5);
\node[below] at (-2,-0.5) {$\mathbf{{e}_1}$};
\node[right] at (-1.5,0.9) {$\mathbf{{e}_2}$};
\end{scope}

\end{tikzpicture}
\vskip 0.4cm
\begin{tikzpicture}[scale=1.3]
\begin{scope}

\pgfmathsetmacro{\cubex}{3.5}
\pgfmathsetmacro{\cubey}{2.8}
\pgfmathsetmacro{\cubez}{0.5}
\draw[ultra thick, black,fill=gray!40!] (0,0,0) -- ++(-\cubex,0,0) -- ++(0,-\cubey,0) -- ++(\cubex,0,0) -- cycle;
\draw[ultra thick, black,fill=gray!90!] (0,0,0) -- ++(0,0,-\cubez) -- ++(0,-\cubey,0) -- ++(0,0,\cubez) -- cycle;
\draw[ultra thick, black,fill=gray!60!] (0,0,0) -- ++(-\cubex,0,0) -- ++(0,0,-\cubez) -- ++(\cubex,0,0) -- cycle;
\end{scope}

\begin{scope}[yshift= -1.5 cm]
\draw[ultra thick, fill=blue!50!, rotate=-15, yshift=-0.35cm,xshift=-0.2cm] (-1.5,0) ellipse (1.25 and 0.5);
\draw[thick, ->] (-1.8,0) to (-.6,1.2);
\node[below] at (-0.5,1) {$\mathbf{J_{dm}}$};
\end{scope}
\begin{scope}[yshift= -1.5 cm]
\draw[ultra thick, ->] (-1.8,0) to (-0.8,0);
\node[below] at (-0.9,-0.1) {$\mathbf{{e}_{W3}}$};
\draw[ultra thick, ->] (-1.8,0) to (-1.8,0.9);
\node[below] at (-2.2,-0.5) {$\mathbf{{e}_{W1}}$};
\draw[ultra thick, ->] (-1.8,0) to (-2.4,-0.5);
\node[right] at (-1.8,0.9) {$\mathbf{{e}_{W2}}$};
\end{scope}
 \end{tikzpicture}
    \caption{ \textbf{Schematic of filament, wall and halo:} 
        \textit{Top panel:} the cylinder represents a typical cosmic filament whose principal axes are given by the $\mathbf{e_1}$, $\mathbf{e_2}$, and $\mathbf{e_3}$ orthogonal vectors, which correspond respectively to the axes of first, intermediate and last collapse. In particular, $\mathbf{e_3}$ gives the filament spine.
        The blue ellipse represents a halo embedded in the filament whose spin points along, $\mathbf{J_{dm}}$.
        \textit{Bottom panel:} shows a similar schematic but for walls. The rectangular cuboid with one edge much smaller than the other two represents a cosmic wall whose preferential axes are given by $\mathbf{e_{W1}}$ (perpendicular to the wall), $\mathbf{e_{W2}}$ and $\mathbf{e_{W3}}$ (within the plane of the wall).
        }
\label{fig:schematicFila}
\end{figure}
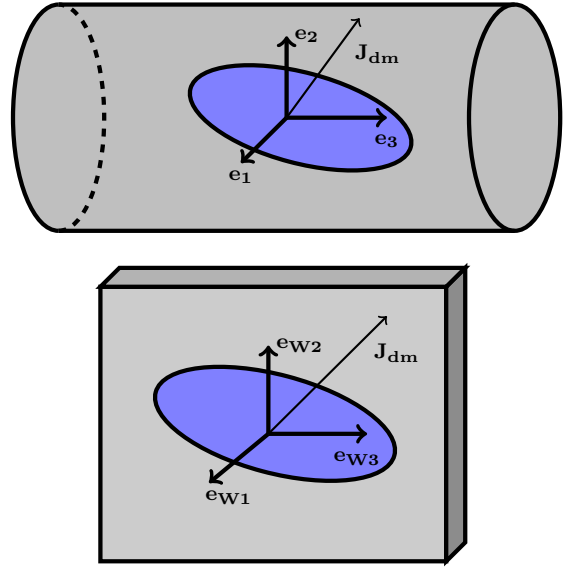

We are calculating alignment angles with respect to all the three preferential axes of the cosmic web. For filaments, we denote the three orientations with $\mathbf{e_1}$, $\mathbf{e_2}$, and $\mathbf{e_3}$. These are given by the eigenvectors of the Hessian matrix of the density field at that location and correspond to the direction of the first, intermediate and the last collapse, respectively.
The principal axes and their configuration with respect to the mass distribution along the filament are highlighted in the top panel of \autoref{fig:schematicFila}. Similarly for walls, we calculate the alignment with $\mathbf{e_{W1}}$, $\mathbf{e_{W2}}$ and $\mathbf{e_{W3}}$ (we use an additional $W$ subscript to distinguish the walls from filaments). The wall principal axes also correspond to the three collapse direction, first, intermediate, and last, respectively, and are illustrated in the bottom panel of \autoref{fig:schematicFila}.

\section{Spin alignment evolution}
\label{sec:filamentAlignment}
\begin{figure*}
    \subfloat{\includegraphics[width = 0.97\columnwidth]{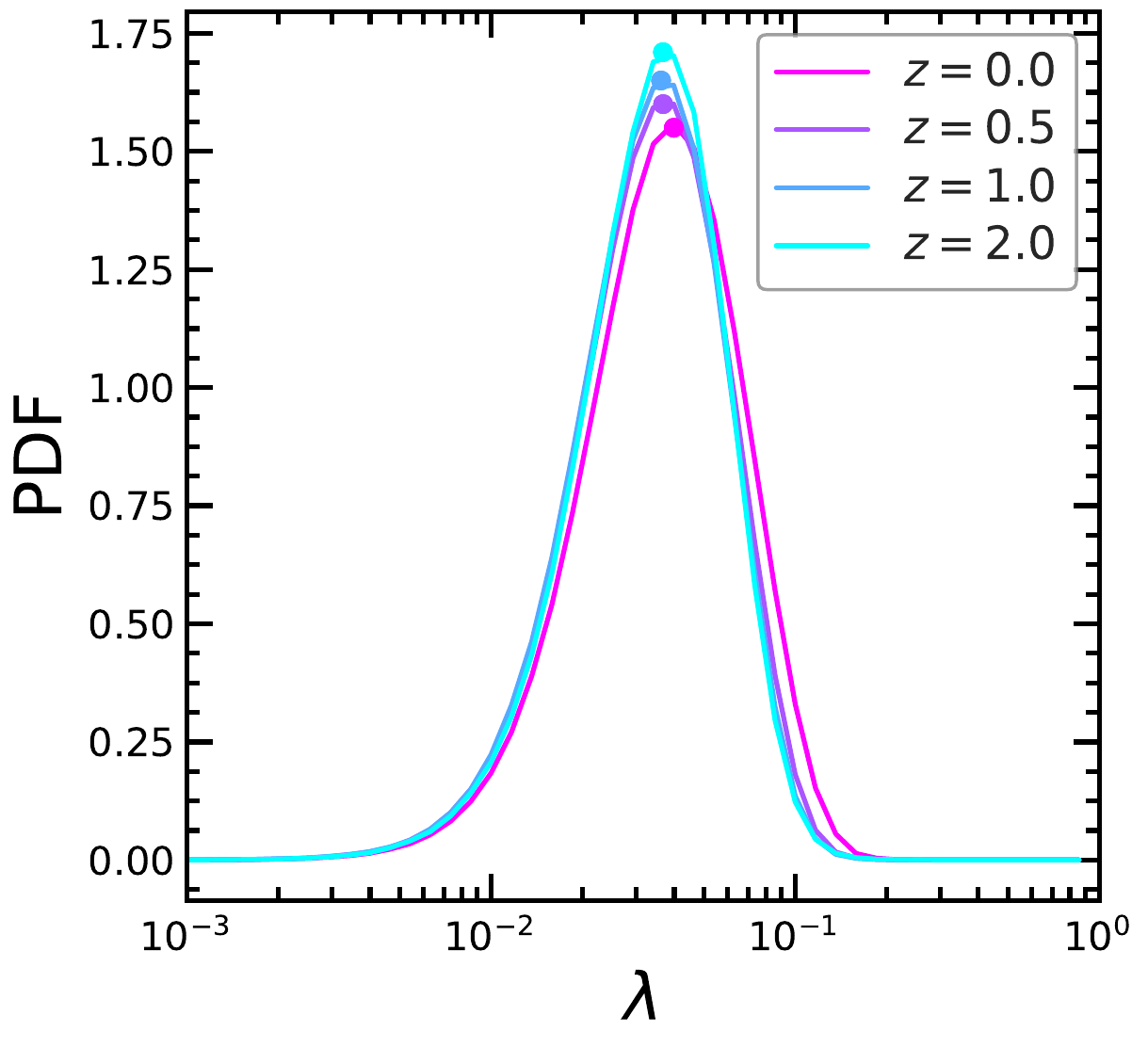}}
    \subfloat{\includegraphics[width =\columnwidth]{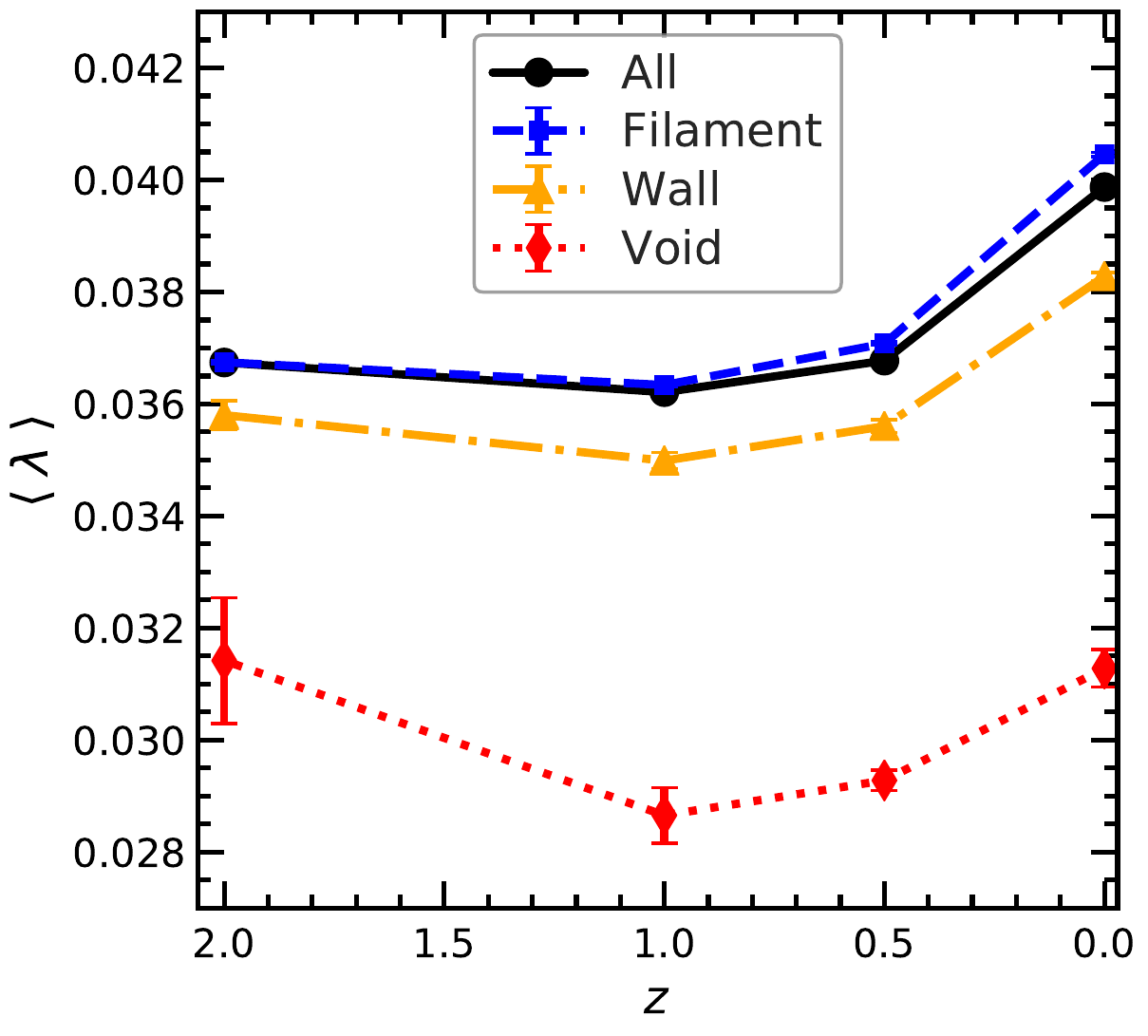}}
    \caption{\textbf{The distribution of halo spins.} \textit{Left panel}: shows the distribution of Bullock spin parameter for all haloes in the mass range $3\times 10^{11}$ to $5\times 10^{12}$\massUnit{} at different redshifts. \textit{Right panel}: shows the median spin parameter for the same haloes as a function of redshift (black solid line). The coloured lines show the median spin for haloes segregated according to their web environment.
    It shows that halo spin acquisition depends on the environment where a halo is located, with haloes spinning the fastest in filaments and walls.
    }
    \label{fig:bullockSpin}
\end{figure*}

We first present an overview of the distribution of halo spins at several redshifts, which is shown in the left-hand panel of \autoref{fig:bullockSpin}. At a each redshift, there is a distribution of halo spins that is well described by a log-normal distribution \citep[not shown here, e.g. see][]{bullock2001, bett2007}. 
The distribution of halo spins shows very small variations between different redshifts indicating that the average halo spin does not vary much with time.
In the right-hand panel of \autoref{fig:bullockSpin}, we plot the time evolution of the median spin parameter for haloes segregated into filaments, walls and voids.
Haloes residing in filaments consistently have a higher spin at all redshifts, followed closely by wall haloes, while voids have systematically lower rotation support. This is in accordance with the results in \citet[][]{GVeena2018} for redshift 0. This is also in agreement with the findings of \citet{hahn2007b} who also have shown, using a different web finder, that filament haloes have a higher median spin at all times.

The dependence of halo spin on environment is the result of two processes. Firstly, within the TTT framework, the halo spin arises from the misalignment between the initial tidal field and the proto-halo mass distribution \citep[e.g.][]{white1984, lee2000, porciani2002,porciani2002TTT2}. The misalignment angle, the strength and anisotropy of the tidal field, and the ellipticity of the initial proto-halo can depend on environment \citep[e.g.][]{codis2015} and would naturally lead to a variation of the halo spin with the web environment. 
This potentially explains why haloes in filaments, which mostly correspond to the regions with strong tidal fields \citep{van_Haarlem1993}, have higher spins than their void counterparts.
Secondly, the deviations of halo spin growth from the TTT predictions depend on the environment, with haloes in higher density regions experiencing a lower growth than those in less overcrowded environments \citep{Lopez2019},
and could explain why we find only a modest difference in median spin between filaments and wall environments.

\subsection{Evolution of spin alignments in filaments}

\begin{figure*}
    \includegraphics[width=\textwidth]{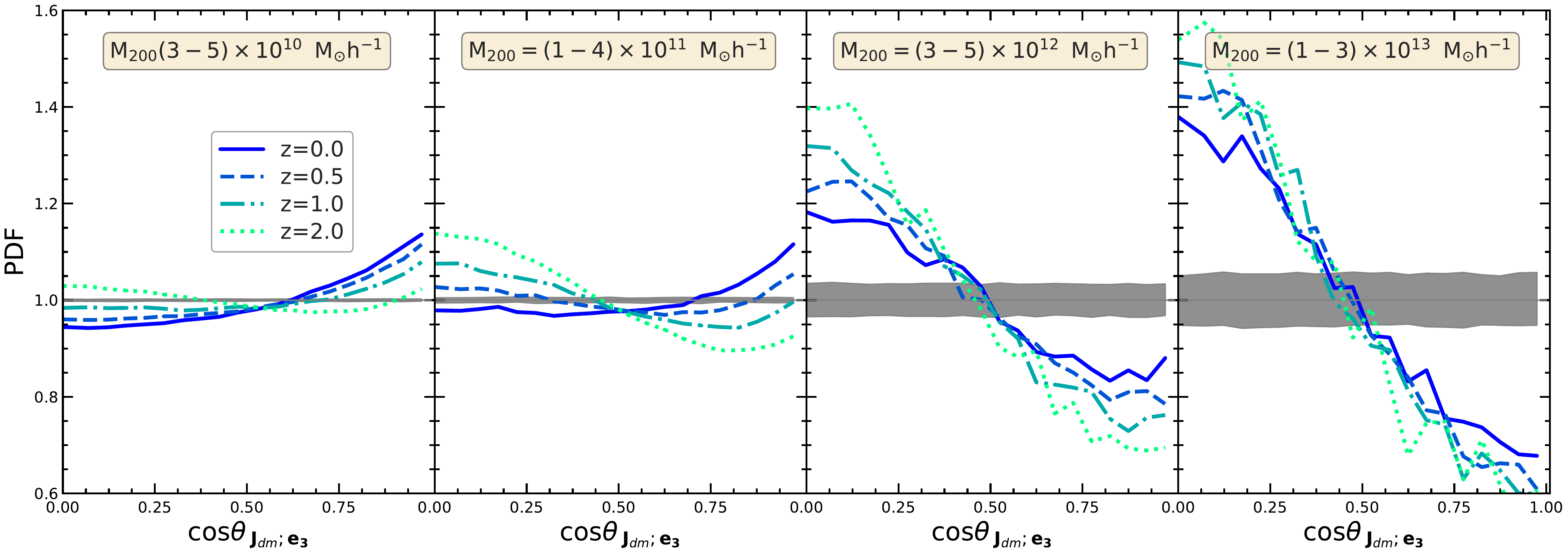}
    \vskip -.1cm
    \caption{\textbf{PDF of the halo spin--filament alignment angle:} 
   Each panel shows the distribution of the spin--filament alignment angle for haloes in different mass bins, with halo mass increasing from left to right (see upper text label in each panel). Coloured lines correspond to different redshifts (see the legend in the left-most panel). The grey horizontal line and its associated shaded region show respectively the mean expectation and the 68 percentile confidence interval when no alignment is expected. Low-mass haloes have an excess of parallel orientations (i.e. PDF is highest at $\cos\theta=1$), while high-mass ones have a propensity for perpendicular orientations (i.e. the PDF is highest at $\cos\theta=0$). The transition halo mass between the two regimes varies with redshift.}
\label{fig:pdfAlignment}
\end{figure*}

In \autoref{fig:pdfAlignment}, we show the time evolution of the angle between halo angular momentum and filament axis. We plot its probability distribution function (PDF) for four mass bins 
(each shown in a different panel) and for the four redshifts studied here. 
Although the alignment angle has a wide distribution of all possible $ \cos \theta $ values, i.e. from 0 to 1 (which corresponds to $\theta = 0 \; \rm{to}\; 90$), it is significantly different from what is expected for a random or isotropic distribution. 
For example, the haloes in the lowest-mass bin show a preferential parallel alignment, that is an excess of spins with $\cos\theta\simeq 1$, which is very low at high redshift and increases towards present day. The second panel, for haloes in the mass range $(1-4) \times 10^{11}$\massUnit{}, neatly illustrates the time evolution of the spin--filament alignment: a propensity for perpendicular configurations at $z=2$ that transforms to an excess of parallel configurations at $z=0$. The highest-mass haloes, shown in the two right-most panels, have preferentially perpendicular spins at all times, although this excess decreases slightly with time.

\begin{figure*}
    \centering
    \includegraphics[width=\textwidth]{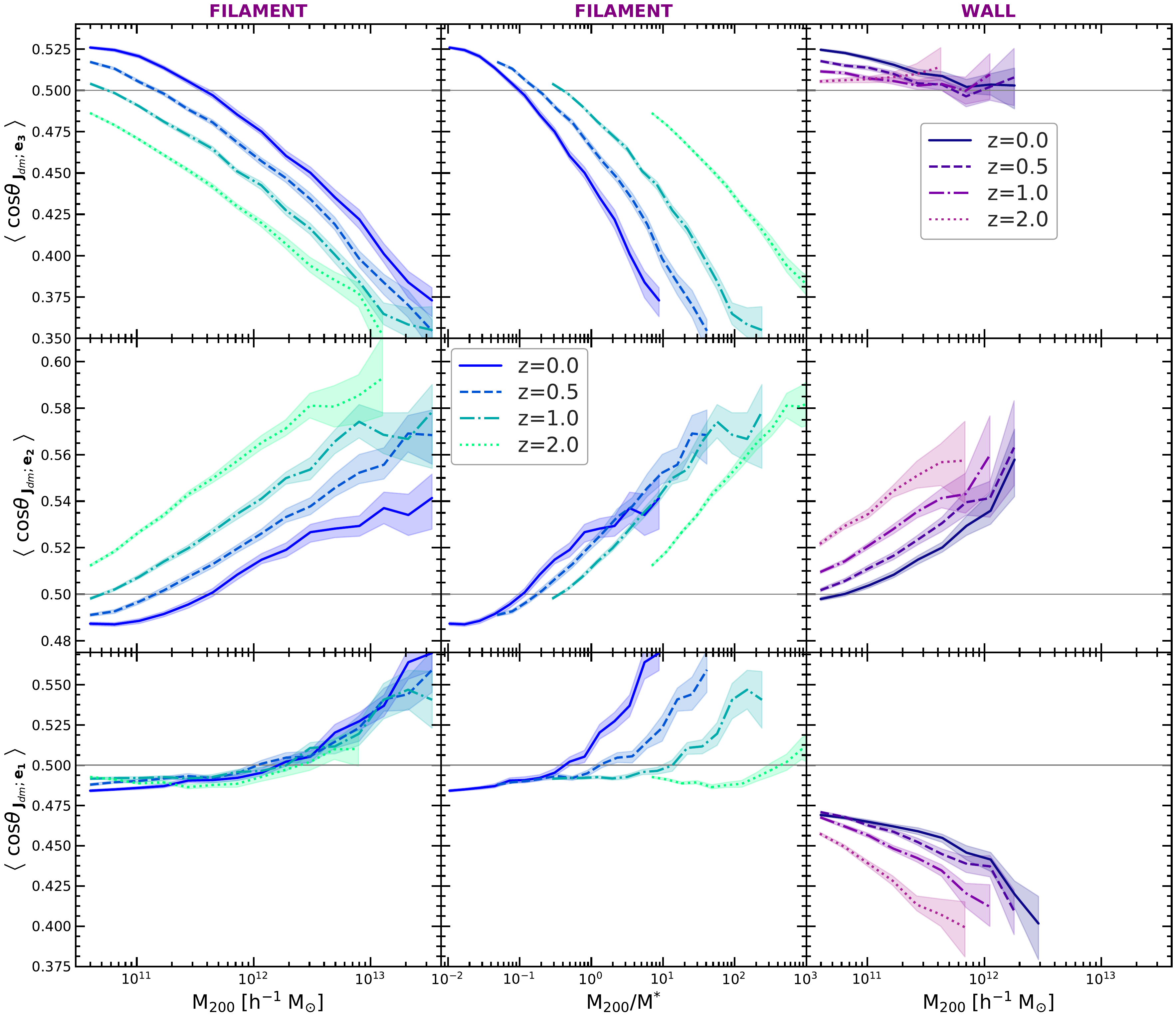}
    \vskip -.2cm
    \caption{ \textbf{The evolution of halo spin alignment in filaments and walls:} 
    It shows the median alignment angle, $\cos\theta$, between the spins of haloes and the principal axes of their environment. The median angle is plotted as a function of halo mass (left- and right-hand columns) and halo mass normalised by the characteristic mass, $M^\star$, at each redshift (middle column). The first two columns are for filament haloes and the right-hand column is for wall haloes. The rows correspond to three environment principal axes: : \eg{} is the axis of last collapse (top row), \et{} is the axis of second collapse (middle row), and \eo{} is the axis of first collapse (bottom row).
    The various colours and linestyles represent haloes at different redshifts, and the shaded region indicates the $1\sigma$ uncertainty when determining the median alignment angle.
    }
    \label{fig:medianAlignment}
\end{figure*}
 
The large number of haloes in our sample allows us to measure very precisely the PDF of the halo spin alignment. This reveals a very interesting find, that is most easily visible in the $z\geq 0.5$ curves shown in the second panel of \autoref{fig:pdfAlignment}. Those PDF show a weak, but statistically robust \revised{(compare with the grey shaded region around the $y=1$ line  which shows the 68\% confidence interval of our calculation)} bi-modality: there is an excess of haloes with $\cos\theta<0.2$ and also an excess of haloes with $\cos\theta>0.8$ (at least compared to haloes with $\cos\theta \sim 0.7$). To our knowledge, this is the first time a bi-modality in the alignment angle has been \revised{ pointed out.} It suggests that there are at least two processes (or classes of processes) that affect the evolution of the halo spin--filament alignment. The first class of phenomena generates preferentially perpendicular alignment, i.e. $\cos\theta \sim 0$, while the second one produces mainly parallel alignments. In general, one of the two effects is dominant, such as for low-mass haloes at $z\leq1$ (left-most panel in \autoref{fig:pdfAlignment}) or for high-mass haloes at all redshifts (right-most panel in \autoref{fig:pdfAlignment}), and no obvious bi-modality can be observed. However, this bi-modality is easily seen for the halo sample that is in the process of changing alignment from preferentially perpendicular to preferentially parallel, when the two classes have a roughly equal impact on the alignment of halo spins.

As we have seen in \autoref{fig:pdfAlignment}, the spin--filament alignment varies with halo mass. To more clearly illustrate this dependence, we study in the left-hand panels of \autoref{fig:medianAlignment} the median alignment angle and its time evolution. We study the spin alignment not only with the filament spine, denoted with \eg{}, but also with the other two principal axes: \et{} and \eo{}.

The top-left panel of the figure shows the alignment with the filament spine, \eg{}, and clearly illustrates that this alignment varies with both redshift and halo mass. A higher fraction of haloes have perpendicular spin orientations at: i) high redshift when comparing equal mass objects, or ii) at higher halo masses when comparing objects at the same redshift. In particular, for most redshifts we find a transition from an excess of perpendicular alignments for massive haloes to a propensity for parallel alignments at low halo masses. The mass at which this transition takes place decreases with redshift. For $z=2$, this transition probably takes places at halo masses below $3\times10^{10}$\massUnit{}, which is the lowest halo mass well resolved by our simulation.

The middle-left panel of \autoref{fig:medianAlignment} shows the spin alignment with the axis of second collapse, \et{}, for filament haloes. Here we find an excess of parallel spin alignments, that increases with halo mass. In particular, we find a transition from parallel alignment at high masses to perpendicular alignment at low masses, with the transition mass being very close to the transition mass found for the alignment with the \eg{} filament axis.
The bottom-left panel of \autoref{fig:medianAlignment} shows the alignment with \eo{}, which is the direction of first collapse. This is rather interesting since it shows hardly any evolution with redshift, although we do find a trend with halo mass that is weaker than the one found for \eg{} or \et{}.

Thus, at fixed mass, the angular momentum of haloes is changing only in the $\mathbf{e_3} - \mathbf{e_2}$ plane and stays roughly at the same angle with respect to $\mathbf{e_1}$, 
which corresponds to a precession of the spin around the \eo{} axis. However, we note that the mass of a halo increases with time, so when comparing equal mass haloes at different redshifts we are not comparing the same objects at different epochs. This observation suggests that the spin--\eo{} alignment varies as the halo grows but in such a way that the spin alignment of the resulting halo is on average the same as for an equal mass halo at an earlier redshift.

The halo mass growth with redshift can be accounted for by normalizing the halo masses by the characteristic mass, $M^\star$, at each redshift.
In a certain sense, this is equivalent to following the growth of haloes relative to the typical halo at a given time.
In the middle column of \autoref{fig:medianAlignment}, we show the evolution of the median alignment after this scaling. If the spin alignment evolution was purely a consequence of the halo mass growth, this should have shifted the curves at different redshifts to overlap each other, however this is not the case. At fixed $M_{200}/M^\star$ values, we find that the halo spin evolves towards a more perpendicular alignment with the \eg{} axis at late times \citep[see also][]{trowland2013, Wang2018a}. This is to be expected since at early times proto-halo spins are predominantly aligned with the \eg{} principal axis of the tidal field \citep{porciani2002,Lopez2020a}, and thus nonlinear spin acquisition (i.e. that is not captured by TTT) leads to the halo spins reorienting themselves such that they are more likely to be perpendicular to the filament spine, \eg{}. Interestingly, at fixed $M_{200}/M^\star$ values, the spin alignment with the \et{} axis hardly changes with time, especially for $z\leq1.0$. This suggests that at late times the spin reorientation proceeds on average as a precession around the \et{} filament axis.

\subsection{Evolution of spin alignments for wall haloes}
We now study the evolution of halo spin alignments with the walls of the cosmic web. The anisotropy and strength of tidal fields in walls are different from those in filaments and we expect deviations from what we have found for filaments. The right-hand column of \autoref{fig:medianAlignment} shows the median spin alignment for haloes in walls, where the rows correspond to the alignment with the wall principal axes: $\mathbf{e_{W3}}$, $\mathbf{e_{W2}}$, and $\mathbf{e_{W1}}$.
 
The schematic of the three preferential axes of walls is given in \autoref{fig:schematicFila}, where $\mathbf{e_{W1}}$ is the axis perpendicular to the plane of the wall, and $\mathbf{e_{W2}}$ and $\mathbf{e_{W3}}$ are along the plane of the wall.

\begin{figure*}
\includegraphics[width=.97\textwidth]{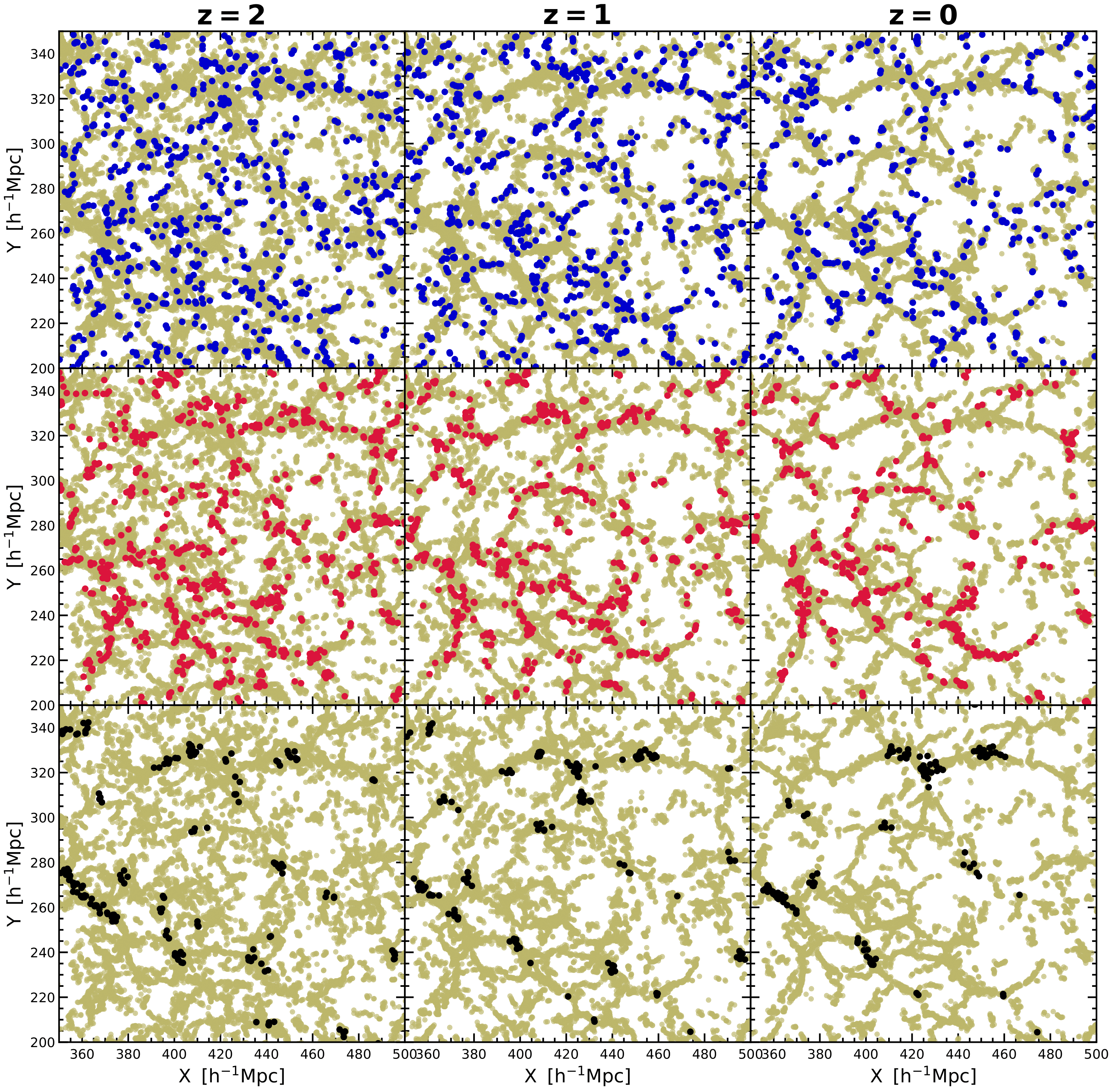}
    \vskip -.2cm
    \caption{\textbf{Haloes in thin, medium, and thick filaments:} \textit{The top row:} shows haloes in thin filaments   ($D_{\rm{filament}} <2$ \distUnit{}) as blue dots and all the filament haloes in that slice as light green dots. From left to right, the panels show the same slice at $z=2$, 1 and 0, respectively. \textit{The center row:} shows the haloes in medium thickness filaments ($D_{\rm{filament}} \in[2,4]$ \distUnit{}) as red dots.  \textit{The bottom row} shows haloes in thick filaments ($D_{\rm{filament}} >4$ \distUnit{}) as black dots. 
    On average, haloes in thin filaments (top row) are at the periphery of the filamentary network bordering the void regions, whereas haloes in thick filaments (bottom row) are mostly at the intersections of massive filaments.
    All panels show the same slice (at different redshifts) which has a comoving thickness of 4.5 \distUnit{}. The x- and y-coordinates are also given in comoving units.
    }
\label{fig:HaloesfilamentThickness}
\end{figure*}

\begin{figure*}
    \includegraphics[width=\textwidth]{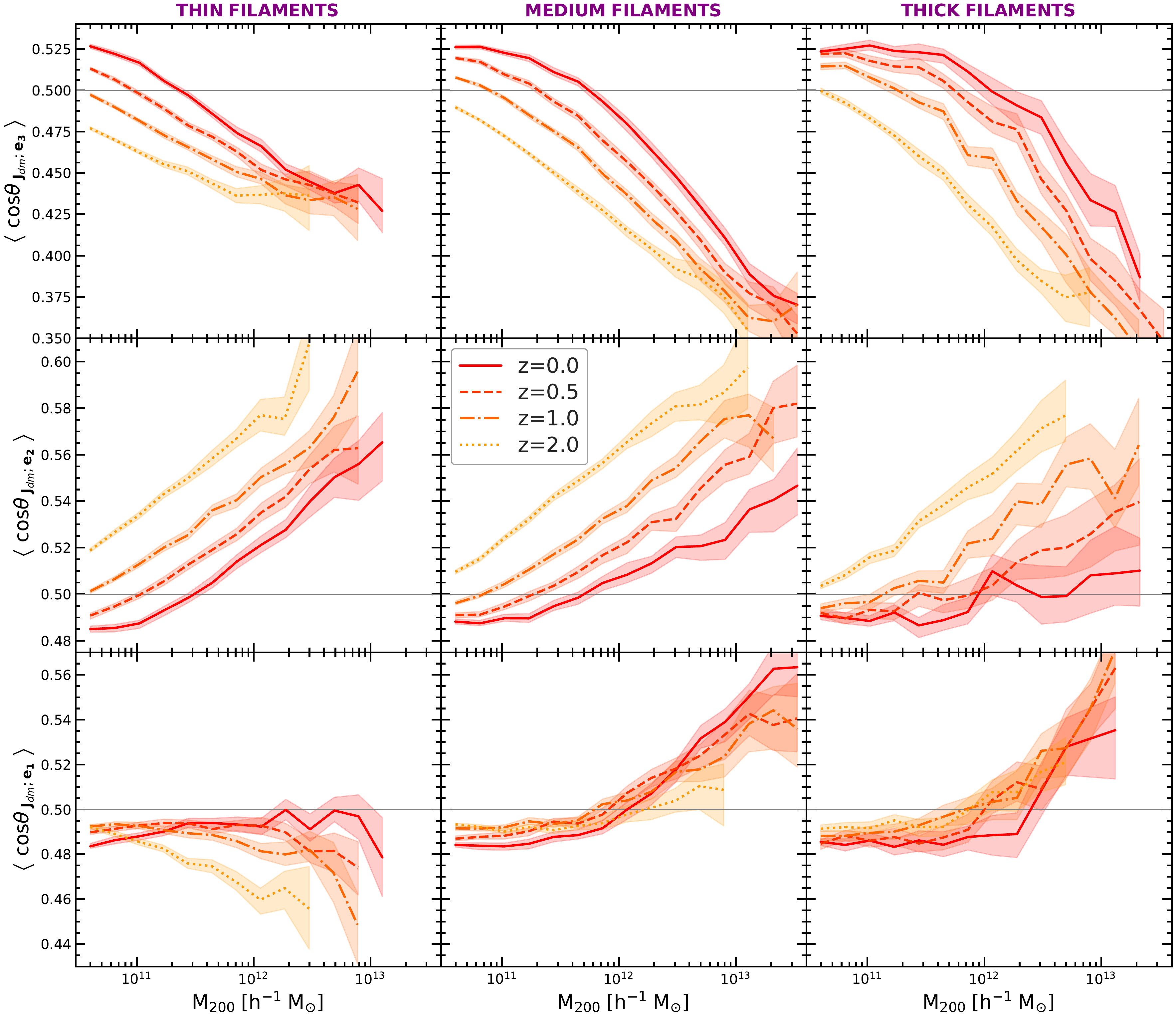}
    \vskip -.1cm
    \caption{\textbf{Spin--filament alignment and its dependence on filament thickness:} 
    The columns show the median alignment between halo spin and filament axes for three different filament subsamples: thin ($D_{\rm{filament}} <2$ \distUnit{}; left column), medium ($D_{\rm{filament}} \in[2,4]$ \distUnit{}; middle column), and thick ($D_{\rm{filament}} >4$ \distUnit{}; right column)). 
    The rows show the alignment with the filament preferential axes (from top to bottom): \eg{}, \et{}, and \eo{}. This plot highlights that the spin-filament alignment and its redshift evolution depends on filament thickness.}
    \label{fig:medianAlignmentFilaThick}
\end{figure*}

The spin alignment of wall haloes is different from that of filament haloes, especially for the $\mathbf{e_{W3}}$ and $\mathbf{e_{W1}}$ axes. With respect to $\mathbf{e_{W3}}$, we find that the alignment is only weakly varying with halo mass and that haloes of all masses and at all redshifts are oriented preferentially along $\mathbf{e_{W3}}$. In particular, we do not find the transition from parallel to perpendicular alignment seen for filaments. In contrast, the alignment with $\mathbf{e_{W2}}$ depends on both halo mass and redshift, and it is nearly identical to that observed for filaments, except that the median $\cos\theta$ is slightly larger for walls than for filaments when compared at equal halo mass and redshift.

The spin alignment with $\mathbf{e_{W1}}$, which is the normal to the plane of the walls, is the most interesting and the one that shows the largest contrast with respect to the filament haloes. On average, wall haloes of all masses have their spin perpendicular to $\mathbf{e_{W1}}$, i.e. the spin is oriented predominantly within the wall plane. The excess of perpendicular configurations is largest at early times and more so for massive haloes. In contrast, most filament haloes have their spins along $\mathbf{e_{1}}$, especially for higher mass haloes.

\section{Filament thickness and spin alignments}
\label{sec:filamentThickness}
The tidal field is responsible for the formation of the cosmic web and the growth of halo spin. Thus, the characteristics of the tidal field, such as its strength and degree of anisotropy, are expected to correlate with the properties of filaments and that of halo and galaxy spins. This correlation manifests itself as a dependence of spins on the nature of filaments, as pointed out by \citet{GVeena2018} who have shown that spin alignments vary with filament thickness. Moreover, the environment of a halo affects the spin by determining the amount of matter and the anisotropic direction along which haloes and galaxies grow, thus potentially further enhancing the correlation between spins and the properties of the web element.

Here, we study the time dependence of spin alignments on the filament thickness in which the halo resides, which up to now has been studied only at $z=0$ \citep{aragon2014,GVeena2018}. This begs the questions: Is the dependence of spin orientation on filament properties due to the recent non-linear spin growth or is it already imprinted in the initial conditions and thus predicted by TTT?

\begin{figure*}                  
    \subfloat{\includegraphics[width=0.5\textwidth]{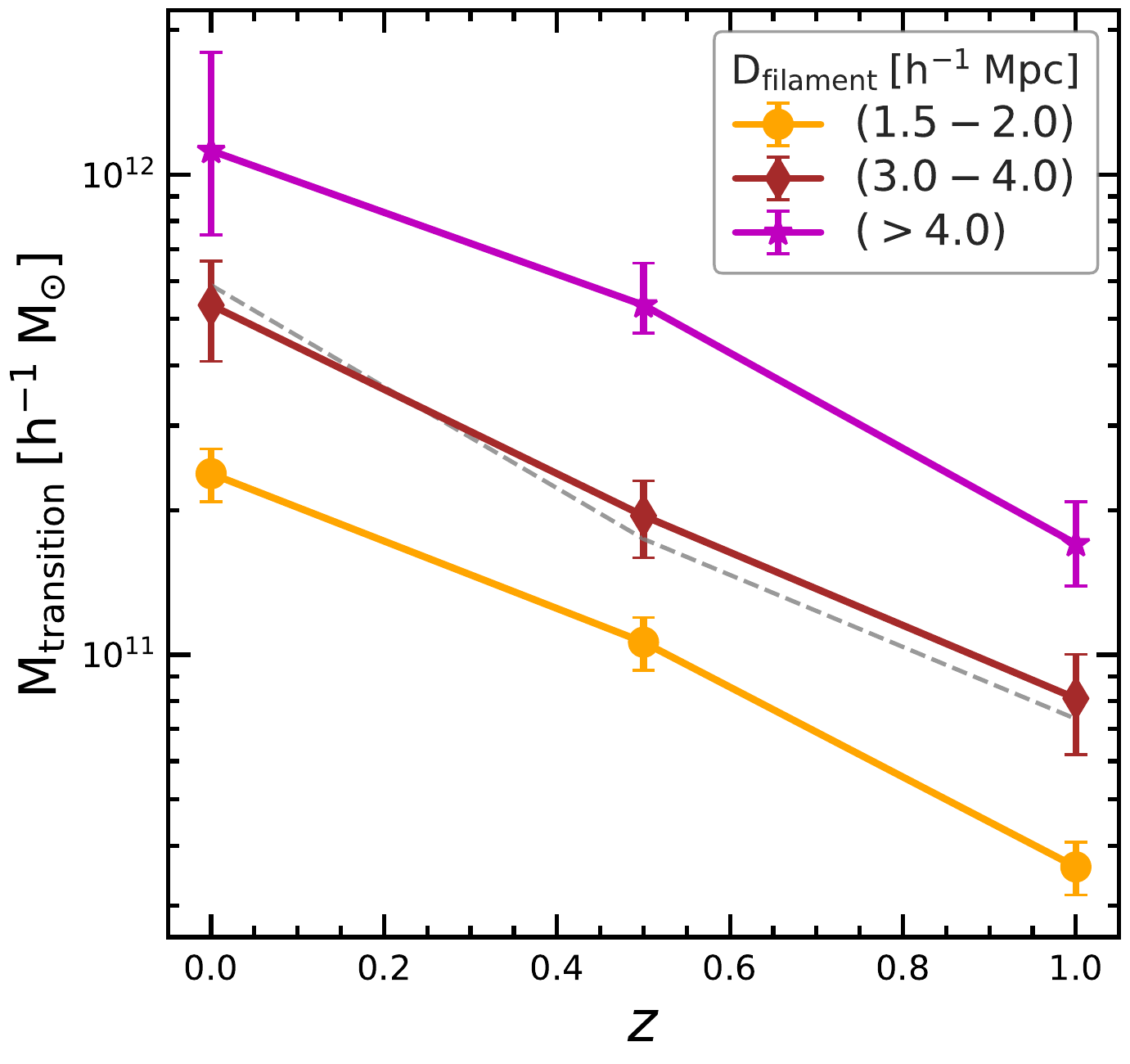}}
    \subfloat{\includegraphics[width=0.5\textwidth]{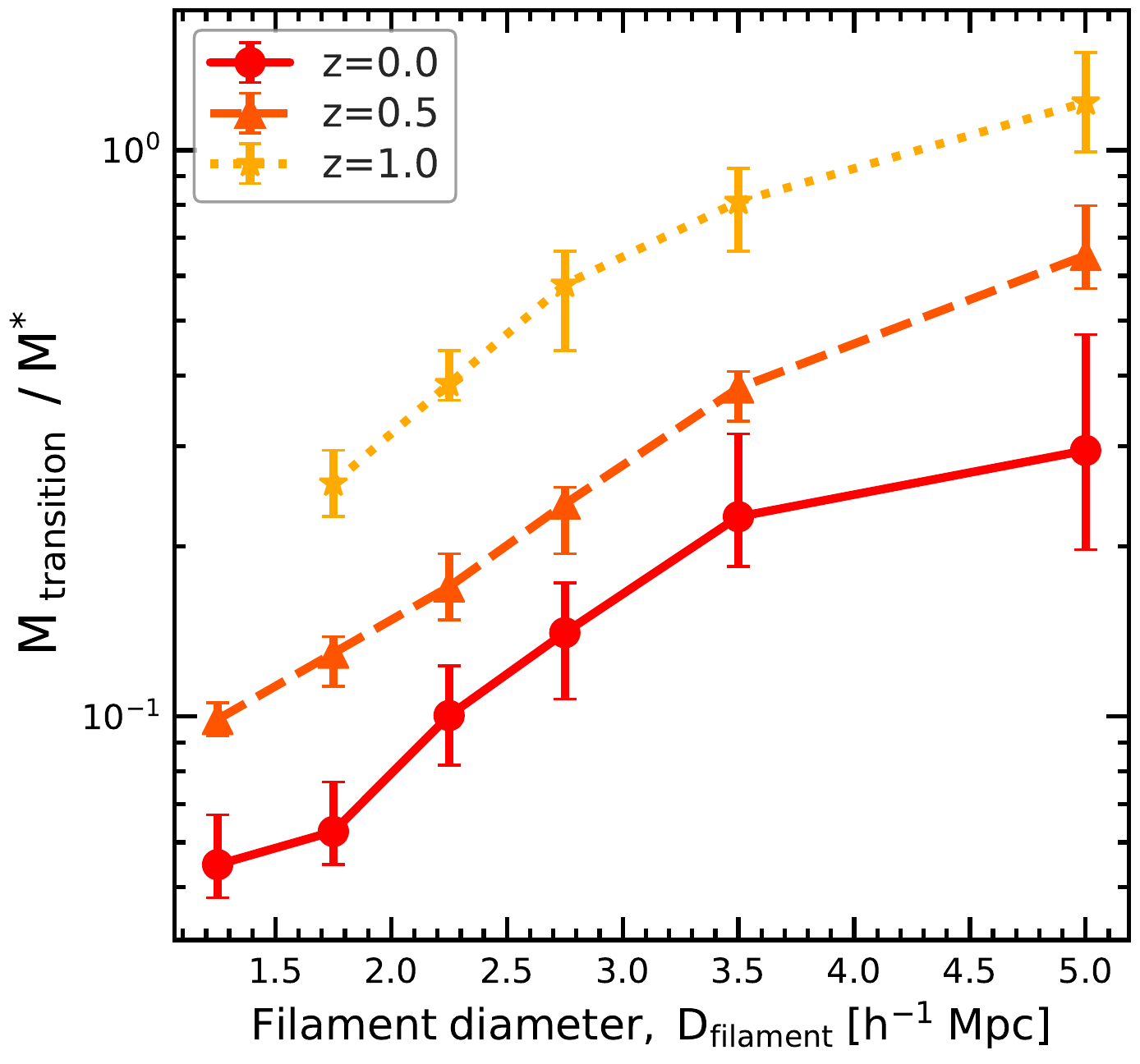}}
    \caption{\textbf{Transition mass and filament diameter:} \textit{The left panel:} shows the redshift evolution of the spin alignment transition mass for filaments of different thickness (see plot legend). \textit{The right panel:} shows transition mass normalised by characteristic mass, $M^{*}$, as a function of filament diameter. The gray dashed line shows the curve $(1+z)^{-3}$. The transition mass increases with filament diameter and also evolves with time.
    }
\label{fig:transitionMassFilaDia}
\end{figure*}
We calculate the filament thickness using the \cite{cautun2014} approach, which represents a local estimate of the filament diameter, which we denote with $D_{\rm{filament}}$. The thickness is obtained by first determining the filament spine, and then by calculating the cross-section centred on the spine needed to enclose all the filament volume elements (i.e. grid cells used for determining the \nexus{} web).

Following this, we define three sub-samples: \textbf{thick filaments} (with $D_{\rm{filament}}> 4$ \distUnit{}), \textbf{medium filaments} (with $D_{\rm{filament}} \in[2,4]$ \distUnit{}) and \textbf{thin filaments} (with $D_{\rm{filament}} <2$ \distUnit{}). Properties of these three filament types, such as linear density and tangential mass profile are studied in detail in \cite{cautun2014}.

\subsection{Halo distribution}
In \autoref{fig:HaloesfilamentThickness} we show the physical distribution of haloes in the three filament sub-samples we just defined. The top-most panel shows haloes in thin filaments as blue dots and all the filament haloes of that slice as light green dots. Panels from left to right correspond to redshift 2, 1 and 0, respectively, and show the time evolution. In the central and lower panels, red and black dots show haloes in medium and thick filaments, respectively. 

The three categories of haloes populate different regions of the filamentary network. Haloes in thin filaments are in the peripheral regions, bordering the main filamentary network. Some haloes are even located inside the much thinner filamentary fabric within voids. Haloes in the medium filaments populate the main arteries of the filamentary network. Haloes in thick filaments shown in the lowest panel are usually closer to clusters, at the intersections of 
prominent filaments. 

Since the haloes are located at distinct locations in the filamentary network, we expect them to have varied dynamical histories. For instance, haloes in the very thin filaments that are part of void regions are isolated from many dynamical processes such as major mergers. Hence, we expect them to retain the original tidal torque acquired during turn-around. At the intersections of the filamentary networks we expect the opposite, i.e, more mergers and accretion along many directions, whereas in the main arteries, we expect a more coherent transfer of mass and angular momentum onto the haloes. We expect these differences to manifest in their angular momentum growth, specifically in the eventual orientation of angular momentum.  

\subsection{Evolution of halo spin alignment and filament thickness}
We now proceed to study the evolution of the spin-filament alignment for the three filament sub-samples we defined at the beginning of this section. The results are shown in \autoref{fig:medianAlignmentFilaThick}, where each row corresponds to the alignment with the preferential filament axes, \eg{}, \et{}, and \eo{}, and each column corresponds to thin, medium and thick filaments, respectively.

\autoref{fig:medianAlignmentFilaThick} illustrates that the spin--filament alignment depends on the filament thickness and that this variation is seen at all redshifts. The size of the difference varies with halo mass and redshift. For example, for ${\sim}10^{12}$ \massUnit{} haloes at $z=0$ we find that $\cos\theta_{\mathbf{J;e_3}}$ is lower for the thin filaments than for the thick ones. This means that haloes of that mass are more likely to have their spins perpendicular to the filament spine if they reside in thin filaments compared to if they would be in thick filaments. For the same ${\sim}10^{12}$ \massUnit{} haloes at $z=0$ we also find that $\cos\theta_{\mathbf{J;e_2}}$ is larger for thin filaments than for thick ones, while $\cos\theta_{\mathbf{J;e_1}}$ shows very little variation with filament thickness.

\begin{table*}
    \centering
    \caption{ \textbf{The values of transition and cross-over masses for 
    various populations of filament haloes.} Transition mass, $M_{tr}$, is the halo mass at which the 
    halo spin--\eg{} alignment changes from preferentially parallel to preferentially perpendicular. The cross-over mass, which we discuss in \autoref{sec:filamentThickness:spin_magnitude}, describes our find that the median halo spin, $\langle\lambda\rangle$, is higher for massive haloes whose spin is perpendicular to their filaments than for those with parallel spins. However, for low-mass haloes the opposite is true, with the cross-over taking place at the mass, $M_+^{\rm{\perp\ vs. \ \parallel}}$. Similarly, high mass haloes in thick filaments have higher spins that those in thin filaments but the opposite is true for low-mass haloes, with the cross-over taking place at halo mass, $M_+^{\rm{thick \ vs. \ thin}}$. We present results for all filament haloes and for haloes residing in filaments of different thickness: thin (with diameter, $D_f<2$ \distUnit{}), medium ($D_f=2-4$ \distUnit{}), and thick ($D_f>4$ \distUnit{}).}
    \label{table:TransitionMass}
	\begin{tabular}{ @{}lcccccc@{}} 
        \hline
		\textbf{Halo population} & \textbf{Redshift} & \textbf{Transition mass} & \textbf{Cross-over mass}   & \textbf{Cross-over mass} \\
         & z & $M_{tr}$  & $M_+^{\rm{\perp\ vs. \ \parallel}}$  & $M_+^{\rm{thick \ vs. \ thin}}$ \\
         & & [ $10^{11}$ \massUnit{} ] & [ $10^{11}$ \massUnit{}] & [ $10^{11}$ \massUnit{} ] \\
        \hline
        
		 \multirow{3}{*}{All filaments} & 0.0 & $3.8 ^{+0.2}_{-0.8}$ & $2.3$ & $3.1$ \\
	    &0.5& $1.5 ^{+0.1}_{-0.3} $ & $1.5$ & $1.0$ \\
    		&1.0 & $0.6 ^{+0.2}_{-0.1} $ & $0.7$ & $0.7$\\ 
    		\hline
				 \multirow{3}{*}{Thin filaments }
				 & 0.0 & $2.3 ^{+0.5}_{-0.1} $ & 1.8 & --\\
	    &0.5  & $0.9 ^{+0.2}_{-0.1}$ & 1.3 & --\\
    		&1.0 &  -- & 0.6 & --\\ 
    		\hline
				 \multirow{3}{*}{Medium filaments} 
				 & 0.0 & $5.4 ^{+1.4}_{-0.1} $ & $2.3$ & --\\
	    &0.5& $2.0 ^{+0.4}_{-0.1} $ & $1.4$ & --\\
    		&1.0 & $0.8 ^{+0.1}_{-}$ & $0.7$ & --\\ 
    		\hline
    						 \multirow{3}{*}{Thick filaments}
    						 & 0.0 & $10 ^{+9.7}_{-1.1} $ & $7.0$ & --\\
	    &$0.5$ & $5.3 ^{+2.3}_{-0.3} $ & $3.0$ & -- \\
    		& $1.0$ & $1.7 ^{+0.7}_{-0.2}$ & $1.1$ & --\\ 
    		\hline
    \end{tabular}
\end{table*}

In terms of redshift dependence, we see an evolution in the alignment with the \eg{} and \et{} filaments axes, and a much weaker evolution in the alignment with the \eo{} axis.
In thin filaments, at redshifts of 2 and 1, we do not find any spin transition from parallel to perpendicular as the majority of the haloes are spinning preferentially perpendicular with respect to \eg{}. The spin transition is seen at later redshifts of 0.5 and 0. In medium and thick filaments, the transition is already seen at a redshift of 1. The fraction of preferentially parallel haloes increases with time and filament thickness at fixed halo mass. Similar observations can be made for the alignment with the \et{} axis, but in this case the fraction of preferentially parallel haloes decreases with time at fixed halo mass.

Compared to \eg{} and \et{}, evolution of the spin alignment with the \eo{} axis  is not very prominent for all three filament sub-samples studied here.
Especially at the low-mass end, there is hardly any time evolution with respect to \eo{}. However, the high mass haloes in thin filaments do show a redshift evolution in their alignment with \eo{}.

To summarise, we observe the dependence of spin--filament alignment on filament thickness at all redshifts studied here. This suggest that this trend, first pointed out in \citet{GVeena2018}, is not due to the recent and highly nonlinear process of spin growth. Instead, it suggests that this difference could have been in place since high redshift and potentially could indicate a systematic variation of the halo spin and its orientation on the local properties of the tidal field (the relevant properties are those that determine the filament thickness). In fact, for low mass haloes ($\lesssim 10^{11}$ \massUnit{}) we find that late time spin growth leads to less variation with filament thickness of the spin-filament alignment (see top and middle rows in \autoref{fig:medianAlignmentFilaThick}).

\subsection{Transition mass: evolution and dependence on filament thickness}
\label{sec:filamentThickness:alignment}

The mass at which the halo population goes from preferentially parallel to perpendicular is known as the transition mass, $M_{\rm{transition}}$. 
We plot in the left panel of \autoref{fig:transitionMassFilaDia} the variation of this transition mass as a function of redshift for a selection of filament diameters. We study only the redshift range $z\leq 1$ since for $z=2$, as we see in \autoref{fig:medianAlignmentFilaThick}, there is no transition from preferentially parallel to perpendicular alignments and hence, we cannot determine this mass (if it exits). This is likely due to the limited mass resolution of our simulation: we only resolve haloes with masses above $3\times10^{10}$ \massUnit{}. It remains to be studied using even higher resolution simulations whether $z=2$ haloes show a spin alignment transition mass.

We find that the transition mass increases towards present-day and also with filament diameter. In particular, filaments of different thickness show the same qualitative behaviour of $M_{\rm{transition}}(z)$, with the only difference being the overall normalisation. It indicates that the evolution of $M_{\rm{transition}}$ is the same for all the three filament samples shown in the figure and that they are different only because they started from a different initial value. 

This raises an important question: Is the evolution in the transition mass due to haloes growing in mass with time? To answer this, we study the transition mass normalised by the characteristic mass, $M^{*}(z)$, at each redshift, which is shown in the right panel of \autoref{fig:transitionMassFilaDia}. There, we show $M_{\rm{transition}}/M^{*}$ as a function of filament thickness, with the three curves now corresponding to different redshifts.
If the increase in transition mass was only because of halo growth, we would've expected the curves at different redshift to coincide when scaled by the characteristic mass. However, this is not the case since $M_{\rm{transition}}/M^{*}$ still changes with redshift. Interestingly, in this case we see a reversal of the trend, the normalized transition mass $M_{\rm{transition}}/M^{*}$ decreases with time.  Therefore,  the evolution of the transition
mass is due to other secondary processes and not only halo mass growth.

\subsection[]{Dependence of halo spin on filament thickness}
\label{sec:filamentThickness:spin_magnitude}

We just have seen that the spin orientation of haloes depends on the filament thickness in which the haloes resides. Could the processes responsible for this trend also lead to systematic variations in the spin magnitude as a function of filament thickness? We explore this question in \autoref{fig:bullockSpinFilaThick}, where
we plot the median spin parameter (see \autoref{sec:analysis}) as a function of mass, for haloes in thin and thick filaments. Note that in this section we are looking at the evolution of the median spin parameter, $\langle\lambda\rangle$, and not alignments. 

At redshift 0 (left-most panel in \autoref{fig:bullockSpinFilaThick}), we find that low-mass haloes in thin filaments have higher spins than their counterparts in thick filaments.

However, for massive haloes this trend is reversed and the spin is higher for haloes in thick filaments. The cross-over between the two regimes takes place at a mass of $3.1\times10^{11}$ \massUnit{}, which we refer to as the cross-over mass, $M_{+}^{\rm{thick \ vs. \ thin}}$. The difference in $\langle\lambda\rangle$ between haloes in thin and thick filaments is small compared to the variance of the spin distribution (see \autoref{fig:bullockSpin}) but it is a robust result (i.e. difference is larger than the uncertainties due to the finite number of haloes; see shaded regions in the figure).

A similar dependence of median halo spin with filament thickness is seen at higher redshift too, as seen in the other three panels of \autoref{fig:bullockSpinFilaThick}. The only difference is that the cross-over mass decreases with increasing redshift to the point that for $z=2$ the cross-over, if any, is outside the mass range available in our simulation (the cross-over at $M{\sim}2\times10^{12}$ \massUnit{} seen for $z=2$ is consistent to noise and likely a spurious effect).

The values of the cross-over mass, $M_{+}^{\rm{thick \ vs. \ thin}}$, at different redshifts are summarised in \autoref{table:TransitionMass} and they indicate that this mass is roughly equal to the spin alignment transition mass at that redshift. This suggests that the same processes that are responsible for the transition in spin--filament orientation are likely to be the ones responsible for the dependence of halo spin magnitude on filament thickness. We will discuss some of these processes in our discussions section, \autoref{sec:discussions}.


\subsection[]{Dependence of halo spin on spin--filament orientation}
\label{sec:filamentThickness:spin_magnitude_vs_alignment}

We now address the final question of this paper: Do parallel and perpendicular haloes have different spin distributions? In otherwords, do haloes spinning preferentially perpendicular to the filament gain angular momentum differently compared to haloes spinning preferentially parallel? 
To explore this, we first classify the subsamples of parallel and perpendicular haloes. \textbf{Parallel haloes} are those whose spin is close to parallel to the spine of their host filaments, i.e. $\cos\theta_{\mathbf{J;e_3}} > 0.8$, which is equivalent to $\theta_{\mathbf{J;e_3}}<\ang{36}$. Similarly, \textbf{perpendicular haloes} are those with $\cos\theta_{\mathbf{J;e_3}} < 0.2$, which corresponds to $\theta_{\mathbf{J;e_3}}>\ang{78}$.

\begin{figure*}
    \includegraphics[width =\textwidth]{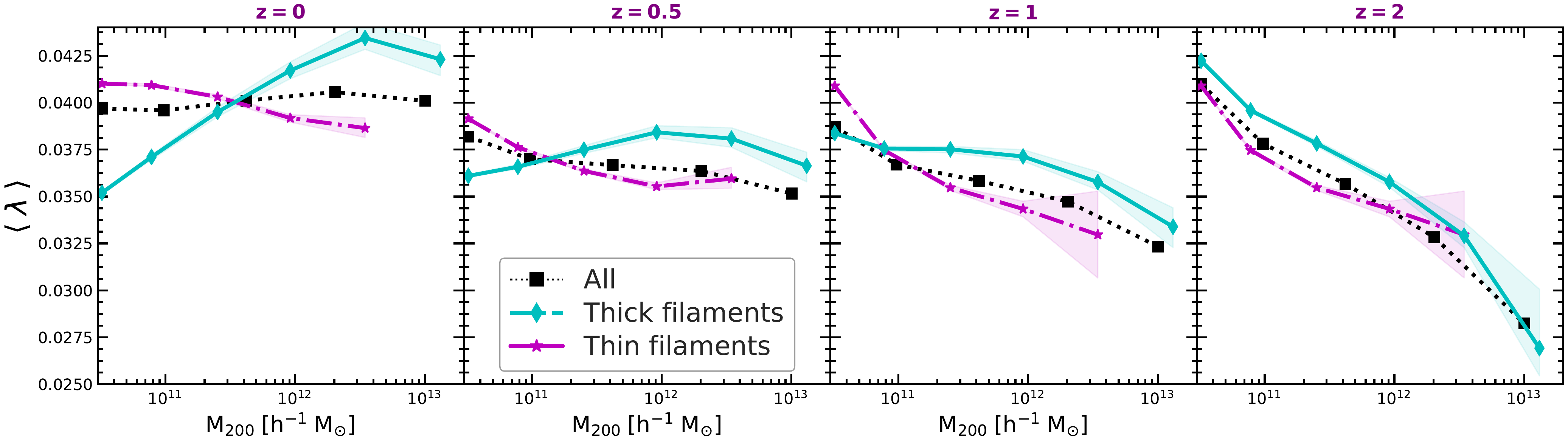} 
    \caption{\textbf{The spin of haloes in thin and thick filaments.} The plot shows the median spin of haloes residing in thin and thick filaments as a function of halo mass. Each panel corresponds to a different redshift (from left to right): $z=0, 0.5, 1,$ and 2.
     At high masses, haloes in thick filaments tend to have a higher median spin than those in thin filaments; for low-mass haloes the trend reverses.
    }
\label{fig:bullockSpinFilaThick}
    \includegraphics[width =\textwidth]{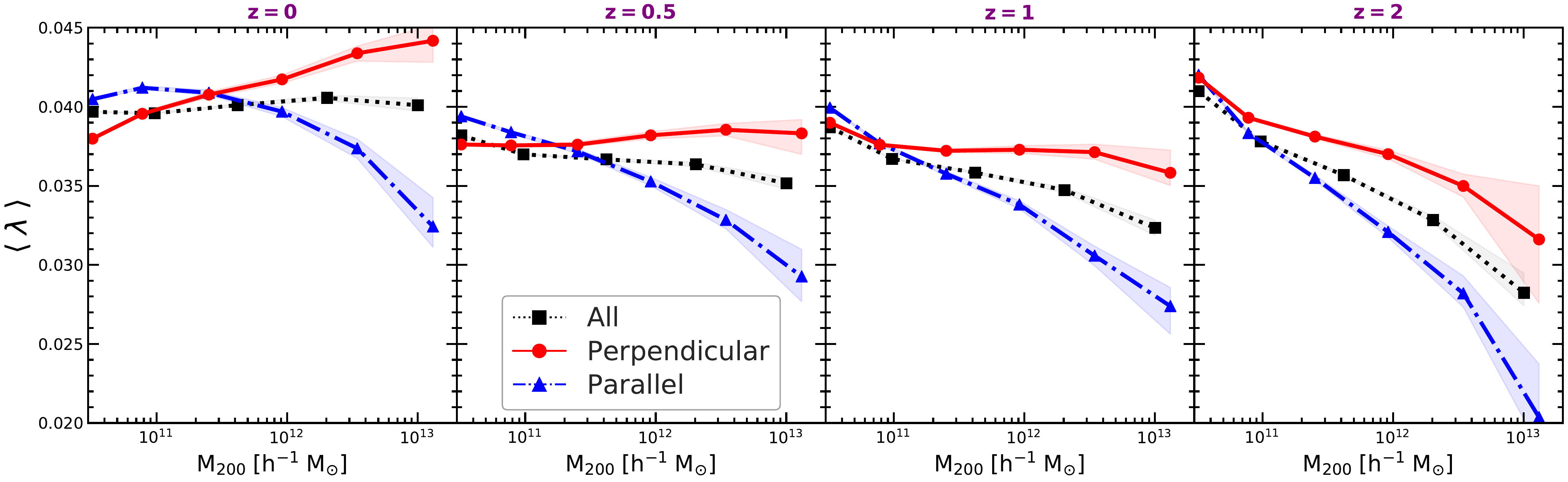}
    \caption{ \textbf{
    The spins of parallel and perpendicular haloes.} 
    We plot the median spin of haloes spinning parallel ($\cos\theta_{\mathbf{J;e_3}} > 0.8$) and perpendicular ($\cos\theta_{\mathbf{J;e_3}} < 0.2$) to the filament spine. As in \autoref{fig:bullockSpinFilaThick}, the panels correspond to different redshifts. At high masses, perpendicular haloes have a higher spin than parallel haloes; the trend is reversed at low masses.
    }
\label{fig:bullockSpinParaPerp}
\end{figure*}

We plot the median spin parameter, $\langle\lambda\rangle$ for these two halo populations as a function of mass at different redshifts in \autoref{fig:bullockSpinParaPerp}. 
At z=0, for haloes less massive than the cross-over mass, $M_+^{\rm{\perp\ vs. \ \parallel}}{\sim}2 \times 10^{11}$\massUnit{}, parallel haloes spin faster than perpendicular haloes, but above this mass, the trend reverses and perpendicular haloes spin faster than parallel haloes. The mass at which this trend reverses decreases as we go to higher redshifts and at z=2, we do not see this crossing over trend.

Interestingly, the cross-over mass seen in the parallel versus perpendicular subsamples and that for thin versus thick filament subsamples are very similar, as can be seen from \autoref{table:TransitionMass}. This could potentially mean one of the effects is a manifestation of the other one, however, this is not the case. Firstly, thin filaments contain slightly more perpendicular haloes than parallel ones (this can be inferred from the top row in \autoref{fig:medianAlignmentFilaThick}, where thin filaments have a larger fraction of perpendicular alignments than thicker ones), however, the dependence of $\langle\lambda\rangle$ in thin filaments is opposite to the dependence for perpendicular haloes. That is, the mass range where $\langle\lambda\rangle$ is larger in thin filaments is the same mass range where $\langle\lambda\rangle$ is larger for parallel, not for perpendicular, haloes. Secondly, when splitting the haloes in thin filaments into parallel and perpendicular subsamples we find the exact same trend as in \autoref{fig:bullockSpinParaPerp}: at the high mass end, perpendicular haloes have a higher spin than parallel one, while the reverse is true for low mass haloes (the same holds true for medium and thick filaments too).

To summarise, the spin magnitude  \revised{is correlated with} both the thickness of the filament in which a halo resides and on the orientation of the spin with respect to the filament spine. These observations underline the multiple complex processes that are responsible for determining the halo spin and its orientation.


\section{Discussion} 
\label{sec:discussions}

In the following, we discuss the most important ramifications of our results and compare them with previous studies in the field.

\subsection[]{The variation of halo spin magnitude with web environment}
\label{sec:discussions:spin_magnitude}

We have found a clear trend between the magnitude of halo spins and the web environment, with the median spin, $\langle\lambda\rangle$, being highest for filament haloes (that is $\langle\lambda\rangle_{\rm{filament}} > \langle\lambda\rangle_{\rm{wall}} > \langle\lambda\rangle_{\rm{void}} $). This trend is present for all the redshifts we have studied ($z\leq 2$) and it does not vary strongly with time. The dependence of halo spin on web environment is a rather small effect (${\sim}10\%$ of the variance of the halo spin distribution) and our very large sample of haloes was essential for revealing this effect. A hint of this trend has been reported in \citet{hahn2007b} but that study lacked the large halo sample needed to robustly quantify the effect. 
The dependence of spin on web environment is another aspect of halo assembly bias, which describes the finding that haloes of same mass but with different properties cluster differently \citep[e.g.][]{Gao2007}. In our case, more clustered haloes, such as those in nodes and filaments, have higher spins than their less clustered equal mass counterparts \citep{Faltenbacher2010}.

The dependence of spin on environment can be ascribed to three potential stages in the growth of haloes. Firstly, it could be a manifestation of correlations present in the initial conditions. In the TTT framework, the spin is due to the misalignment between the shape of protohaloes and the initial tidal field. This misalignment can vary systematically from region to region \citep[e.g.][]{weyedb1996,desjacques2008,rossi2013,codis2015}, and in particular can be different for the regions that will collapse to form filaments, sheets and voids. Secondly, the maximum expansion of a halo and thus the time available for halo spin to grow (within TTT most of the halo spin is acquired at or before maximum expansion of the halo) also depend on environment \citep[e.g.][]{hahn2007,Lopez2019}. Thirdly, the spin growth is affected by nonlinear processes, such as mergers, which can also imprint an environment dependence. For example, \cite{hetznecker2006} have shown that the spin parameter increases considerably for haloes that have undergone major mergers, which are expected to be more common in crowded environments such as filaments. The late-time spin growth is affected by the degree of anisotropic accretion and, in particular, by highly anisotropic infall of satellites \citep[][see \citealt{Tormen1997} for a detailed study of this aspect in the case of clusters]{Libeskind2014,Gonzalez2016,Shao2018}. Our finding that the dependence of spin magnitude on environment is roughly the same since at least redshift, $z=2$, suggests that early time processes, such as the first two we discussed, are likely the most important ones.

\subsection[]{ Evolution of halo spin alignment with the cosmic web}
\label{sec:discussions:spin_alignment}

We have studied the evolution of the alignment between halo spin and the preferential axes of the host filaments and walls. In general, we find that the alignment varies with halo mass and redshift \citep[see also][]{aragon2007,aragonPhd2007,codis2012,trowland2013,Wang2018a}. For example, in present day filaments, the spins of low-mass
haloes are preferentially parallel while those of high-mass haloes are preferentially perpendicular to the filament spine, \eg{}. At fixed halo mass, a higher fraction of haloes have parallel spin--\eg{} alignment at later times, while the opposite is true for the spin--\et{} alignment. For filaments, we find the largest evolution in the spin alignment with the intermediate, \et{}, and last, \eg{}, axes of collapse, and hardly any evolution with \eo{} (but nonetheless the spin--\eo{} alignment varies with halo mass).

To account for halo mass growth, we also have studied the evolution of the spin--filament alignment as a function of the normalized mass,
$M_{200}/M^{*}$, i.e. the halo mass in units of the characteristic collapse mass $M^{*}(z)$ at redshift $z$. In this case, at fixed $M_{200}/M^{*}$,
we find that a higher fraction of haloes have spins perpendicular to \eg{} at later times, while the opposite trend is seen for \eo{}. The spin--\et{} alignment hardly changes with time. This indicates that individual haloes, on average, tend to reorient their spins to be preferentially perpendicular to the filament spine, \eg{}. One explanation for this is the anisotropic accretion of substructures along the host filament spine \citep[e.g.][]{Libeskind2014,Shao2016,Shao2018}, which leads to the halo major axis being oriented along the filament spine and the spin perpendicular to the filament spine \citep{van_Haarlem1993,libeskind2013,GVeena2018}.
    
In walls, we find that the most important spin alignments are with the \eo{} and \et{} preferential axes, and only a weak alignment with \eg{}. In particular, at all halo masses the spins are perpendicular to the normal to the wall, \eo{}, indicating that the halo spins are preferentially pointing in the plane of the wall \citep{aragon2007, aragonPhd2007, aragon2014, Wang2017a, Codis2018}.  At fixed mass, the halo spins reorient such that fewer wall haloes have spins perpendicular to \eo{} at later times, more akin to filament haloes. This fits well with the picture of mass transport across the cosmic web environment in which wall haloes are expected to migrate to filaments \citep{cautun2014,Wang2017a}. \revised{We note that the halo spin alignment in walls shows the largest variation between different studies. This is probably a consequence of the challenges involved in identifying walls, which are rather underdense structures populated by low-mass  haloes and galaxies. For example, \citet{Codis2018} find a halo spin alignment for walls that is similar to our results for thin filaments, potentially indicating that the haloes assigned by DISPERSE (which is the web finder used by \citeauthor{Codis2018}) to walls correspond to haloes mostly identified by \nexus{} as part of thin filaments. \nexus{} finds a large difference in the spin alignment of filaments and wall haloes that suggests that \nexus{} is better at separating between these two cosmic web environments.}

\subsection{Dependence of spin alignments on filament thickness}
\label{sec:discussions:spin_alignment_thickness}

Motivated by the results of \citet{GVeena2018}, who have shown that at $z=0$ the halo spin alignments vary with filament properties, we have studied the evolution of the spin--filament alignment for filaments of different thickness. The alignment shows a pronounced variation with filament diameter, with a higher fraction of equal mass haloes having perpendicular spins if they reside in thin filaments compared to thick filaments. This is present at all the redshifts we have studied (i.e. $z\leq2$) and shows a mild growth with redshift, i.e. the difference is somewhat larger at earlier redshift. This indicates that the dependence of spin--filament alignment on filament properties is set at early times and it is not a late-time effect. One potential explanation is that the dependence on filament thickness is set in the initial condition, that is the misalignment between the moment of inertia of the protohalo and the tidal field is correlated to the size of the $z{\sim}0$ filaments.

The dependence of the spin alignment on filament properties highlights that the tidal environment in which a halo is located influences halo
growth and therefore its eventual angular momentum orientation \citep[see e.g.][]{jainbert1994}. The correlation between tidal field and
halo spins has figured prominently in the theoretical studies of \cite{Lee2020} \citep[also see][]{Lee2019} and
\cite{porciani2002, porciani2002TTT2}. In a recent study, \citet{Wang2018a} showed that haloes in regions with low tidal anisotropy
have spins orientated preferentially parallel to \eg{} (the last collapse axis) while haloes in regions with high tidal anisotropy have
spins preferentially perpendicular to \eg{}. The  dependence of the spin orientation on filament thickness and the effect indicated by
\cite{Wang2018a} are potentially related, but it is unclear to what extent.

To get more insight into the question of the influence of the thickness of filaments, and its relation with the tidal force field, we 
need to identify the factors that determine the strength and thickness of filaments. The cosmic web theory of \cite{bond1996}
points out three major influences that determine a filament's properties. For example, a strong tidal field translate into thicker and more massive filaments \citep{weyedb1996,vdw2008}.  The most prominent filaments tend to form in between galaxy clusters because of the strongly anisotropic
force field induced by such configurations. Stronger tides can be induced by more massive clusters and/or shorter
mutual distances. 

Also, we
know that filaments are not uniform structures, and tend to attain a considerably higher density and diameter at the location where
they connect to the outskirts of clusters \citep{cautun2014}. Indeed, in this study we have found that, in general, the haloes in thick filaments are those that are close to galaxy clusters (see lowest row in \autoref{fig:HaloesfilamentThickness}, also see
Figure 17 in \citeauthor{cautun2014} and \citealt{Galarraga-espinosa2020}). In conclusion, the dependence of spin on filament nature highlight the fact that the strength and the
degree of anisotropy of the tidal field plays a crucial role in determining halo properties, such as spin magnitude and orientation
\citep{bondmyers1996, desjacques2008, codis2015, Paranjape2018}.

\subsection[]{The evolution of transition mass for spin alignment}
\label{sec:discussions:transition_mass}

As we discussed, the spin alignment with the filament spine, \eg{}, changes from a propensity for parallel orientations for low-mass haloes to one of perpendicular configurations at high mass. The halo mass corresponding to this change, known as the transition mass, increases with both time and filament thickness. In particular, the variation with filament thickness is rather large, with more than one order of magnitude variation between the thinnest and thickest filaments \citep{GVeena2018}. Due to the multiscale nature of the method, \nexus{} identifies many more thin filaments than the majority of other web finders \citep{Libeskind2018} and thus determines a systematically lower transition mass than previous studies.

\revised{There are several processes that could be responsible for the  dependence of the transition mass on filament thickness identified in the present study. One suggestion is that of \citeauthor{Laigle2015},
       who linked the transition mass to the size and nature of the multi-flow region, and hence
       also of the scale and thickness of a filament. Additional nonlinear processes connected to the
       filamentary nature of accretion of matter may also play a role, as suggested by a
       recent work of \citet{Lopez2020a}}.

\revised{The transition mass of the full population of filament haloes decreases with redshift as $(1+z)^{-3}$, which in good agreement with the redshift trend found by \citet{codis2012} (see right panel of Figure 3 in \citet{codis2012}) although our values are almost an order of magnitude lower than theirs.}  We have also studied the evolution of the transition mass for filaments of different thickness to find that the relative growth rate is the same in all cases. This is another indication that the dependence of spin alignment on filament thickness is not a late time process but actually is in place before redshift 2.

\subsection{Dependence of spin magnitude on filament properties and spin--filament alignment}
\label{sec:discussions:spin_magnitude_parallel_vs_perpendicular}

In \autoref{sec:discussions:spin_magnitude} we discussed how the magnitude of the halo spin depends on web environment, being highest in filaments. We have also found that the spin magnitude depends on filament properties, such as their thickness. Massive haloes have a slightly higher spin if they reside in thick filaments compared to thin filaments. For low-mass haloes, the opposite relation holds, with haloes in thin filaments having higher spin. The cross-over between the two regimes takes place at a halo mass that is roughly equal with the spin alignment transition mass at that redshift. 

This highlights that the two phenomena, i.e. variation of spin magnitude with filament thickness and spin--filament alignment, are highly correlated and likely due to the same physical processes.

We have also found that the spin magnitude depends on whether a halo is oriented parallel or perpendicular to the spine of its host filament. Similarly to the dependence of spin magnitude on filament thickness, here we also have found a mass dependent trend. High mass haloes spin faster on average if their spins are oriented perpendicular to \eg{}, while the opposite is true for low-mass haloes. The cross-over mass between the two regimes is the same as the cross-over mass for the dependence of the spin magnitude on filament thickness. This might suggest that the two effects are the same, however that is not the case, as we have discussed at length in \autoref{sec:filamentThickness:spin_magnitude_vs_alignment}.

One possible explanation for the dependence of the spin magnitude on filament thickness and spin--filament alignment has to do with the collapse time of a halo. \citet{Lopez2019} have shown that on average haloes that collapse later (their \textit{W}-sample) end up having higher spins than haloes which collapse early. The location of these late collapse haloes depends on their mass. High-mass haloes are more clustered (i.e. in our language they are found in thicker filaments), while low-mass haloes are less clustered (i.e. preferentially found in thin filaments). This offers an elegant explanation for the trend between spin and halo thickness found in this paper. 

The \citeauthor{Lopez2019} (see also \citealt{Lopez2020a}) results also offer an explanation for the trend between spin magnitude and spin--filament alignment. At all masses the late collapsing haloes have spins that are preferentially perpendicular to their host filaments, and explains why we find that massive haloes spin faster if they have perpendicular spin--filaments alignments. 

However, the \citeauthor{Lopez2019} results do not explain the inverted trend we find for low-mass haloes, where the fastest spinning haloes are the ones with parallel spin--filament alignments. The discrepancy could be due to the fact that \citeauthor{Lopez2019} have used a different halo and filaments definition than ours and have studied all haloes while we have analysed only filament haloes. Filaments contain the majority of high-mass haloes (i.e. $M_{200}\in[1,50]\times10^{12}$ \massUnit{}) but only around half of the low-mass ones (i.e. $M_{200}\sim 1\times10^{11}$ \massUnit{}; see \autoref{fig:numberFrac}), so we can only make qualitative comparisons but not draw an explicit connection between the halo sub-samples of the two studies.

The relation between the collapse time of a halo and filament thickness has been explored by \citet{Borzyszkowski2017}. They have found that haloes in filaments that are thin compared to the halo size (i.e. their \textit{accreting} sample) grow by accretion from the filaments surrounding them, which imparts them a tendency for spin orientations perpendicular to the filaments feeding them, and have on average late collapse times. In contrast, the mass distribution of haloes embedded in thick filaments for their size (i.e. \textit{stalled} haloes in the \citeauthor{Borzyszkowski2017} nomenclature) has been set in place since early redshift and recent accretion mostly takes place along directions perpendicular to the host filament spine in which the halo is located. In this case, the accreted mass brings in angular momentum that is preferentially parallel to the filament spine. 

To summarise, haloes that have collapsed sooner are likely to have less spin compared to haloes that collapse later on, which have more time to grow their spin through tidal torques.

\subsection{\revised{The diversity of spin--filament alignments}}

\revised{Our results show that the halo spin orientation depends on filament nature at least for redshift $z<2$, and likely at higher redshift too. Different web finding formalisms concentrate on different physical and structural aspects of filaments. This may result
in somewhat different alignment signals of haloes with respect to the filament population. It suggests that comparisons between the alignment results based on different web finders are not
immediately straightforward, and should take account of
several subtle effects involved to allow meaningful conclusions. Our recent studies \cite{GVeena2018, GVeena2019} have demonstrated how the different results obtained for three different filament population identifications - comprising different versions of the Nexus formalism based on density and velocity field aspects of filaments, as well for the Bisous formalism \citet{Tempel2014a, tempel2016} - could be fully explained and
quantified on the basis of our knowledge of their definition. It shows how a meaningful comparison of filament and
environment dependent properties of haloes and galaxies should include a careful assessment of the subtle differences 
between the physical and structural properties of the filament population. 
This not only concerns aspects such as spin orientation, but also a larger diversity of possibly environment dependent halo and such as density profiles, galaxy and gas content \citep[e.g.][]{cautun2014,Galarraga-espinosa2020,Galarraga-espinosa2020b}.} 


\revised{The \nexus{} method contains two main parameters that determine the resulting filament population. First, there is the set of smoothing scales that are combined to obtain the multiscale cosmic web. In principle, one should use all scales, but in practice there is a minimum scale determined by the grid cell size used for the calculation. Increasing this minimum scale leads to first removing thin filaments, and then, as the scale is raised further, medium filaments. Thus, varying the minimum scale can impact the  alignment signal by removing the thinner filaments. The second parameter is the threshold used to identify robust nodes, filaments, and walls. This is calculated automatically by the method, but in principle could be selected manually too. Raising this threshold above the optimal value has two effects: removes thin filaments, and, at the same time, removes the outer shells of thick filaments.}

\section{Conclusions}
In this paper, we have studied the late time ($z\leq2$) growth of the angular momentum of haloes and how this property is affected by the cosmic web environment in which a halo resides. We have identified the cosmic web elements using the \nexus{} multi-scale algorithm that has been designed to capture the hierarchical and scale-free character of the large-scale web. Our study made use of a large volume and very high resolution N-body simulation, Planck-Millennium, that resolves the formation of dark matter haloes over several decades in halo mass. To ensure that our halo properties are well resolved and converged, we have studied only objects with at least 300 particles, corresponding to a halo mass above $3.2 \times 10^{10} $ \massUnit{}.

We have focused on two factors of halo spin growth: i) how the spin magnitude varies with web environment, and ii) the evolution of the orientation of halo spins with respect to the preferential axes of filaments and walls.
The main conclusions of this study are as follows:
\begin{itemize}
    \item The median spin parameter, $\langle\lambda\rangle$, is highest for haloes in filaments followed by haloes in walls and is least for void haloes. This trend is seen for all redshifts we have studied (the distribution of halo spin parameters hardly changes since $z=2$). 
    
    \item The spins of haloes show a preferential alignment with the spine of their host filaments that depends on halo mass and redshift. Massive haloes show a propensity for perpendicular alignments while low-mass haloes have an excess of parallel alignments.
     The spin transition from perpendicular to parallel orientations is seen for $z\leq 1$ but not at z=2, perhaps due to the limited mass resolution of our simulation. 
     
    \item The mass at which the 
    halo spin--filament alignment changes from preferentially parallel to perpendicular, known as the spin transition mass, evolves with time. It is highest at present and decreases towards higher redshifts. 
    
    \item By scaling the halo mass by the characteristic collapse mass at each redshift, we have shown that the spin of individual haloes reorients such that, on average, it becomes more perpendicular to the filament spine at later times. When expressed in units of the characteristic collapse mass at each redshift, the spin transition mass decreases with time. 
    
    \item The spin--filament alignment varies with host filament properties, such as filament thickness.
   At all redshifts, the fraction of halo spins with perpendicular orientations is largest in thin filaments and decreases in thicker filaments. 
   
   \item Similarly, the transition mass grows rapidly with filament thickness (it varies by an order of magnitude between thin and thick filaments).
   The values and the redshift evolution of the spin transition mass also depends on  filament thickness.
    
    \item The dependence of the spin alignment on filament thickness may not be a late time effect but it is likely already set at high redshift (possibly in the initial conditions). This can be inferred from the fact that the relative growth rate of the spin transition mass is the same for filaments of different thickness. 
    
    \item The magnitude of the halo spin, similarly to its orientation, depends on filament thickness. The spin of massive haloes is higher if they reside in thick filaments compared to thin filaments, while the opposite is true for low-mass haloes. 
    
    \item Similarly, the magnitude of the halo spin depends on the halo spin--filament angle. Massive haloes in which the spin is perpendicular to the filament spine have higher spin than haloes in which the spin is along the filament spine, while the converse is true for low-mass haloes.

\end{itemize} 

Our analysis highlights the complex relation between halo spin and the web element in which a halo resides. Both the magnitude and orientation of the halo spin depends on whether the halo is inside a node, filament, sheet or void, and for filaments, which host the majority of haloes, it also depends on filament properties, such as thickness. Understanding this problem is further complicated by the fact that the correlation between spin and web environment depends also on halo mass, with haloes of different masses showing distinct trends with redshift. 

The implications of our results have been discussed at length in \autoref{sec:discussions}. Here, we would like to highlight that many of the trends we have found, such as the dependence of halo spin magnitude on web environment or the dependence of the spin alignment angle on filament thickness, are mostly set either in the initial conditions or in the early stages ($z>2$) of halo growth. This hypothesis can be investigated by tracing back in time all the dark matter particles associated to a late time halo and determining how the spin of that particle distribution changes in time. This is analogous to the TTT approach in which the spin of the present day halo is given by the integrated effect of the tidal field acting on the particle distribution that ends up collapsing to form the $z=0$ halo. This approach will be applied in the upcoming study of \citet{Lopez2020a}, who will study the halo spin--filament alignments in the context of TTT. A recent study by \cite{Motloch2020} even attempted to follow this approach in an observational context, within the context of the (still limited) dataset of the SAMI or MaNGA IFU galaxy surveys. 

The spin orientation of dark matter haloes is imprinted in the rotation of galaxies \revised{\citep[for a recent review see][]{Somerville2015}} 
and thus it should in principle be possible to study several environmental trends shown in this study using large surveys. Although galaxy spin--filament alignments have been detected in observations \citep[e.g.][]{tempel2013,Welker2020}, there have not yet been any studies that shows trends with redshift or filament properties. Most of the current large surveys, such as SDSS, are limited to low redshift and bright galaxies, which means that they mostly identify only the most prominent filaments. However, future surveys, such as the DESI Bright Galaxy Survey \citep{Smith2019}, will provide observations several magnitudes deeper and will cover a wider redshift range, which will allow for the detection of filaments of various thicknesses (e.g. see the \citealt{Alpaslan2014} study of filamentary tendrils in GAMA) and for the analysis of the galaxy spin--filament alignment at multiple redshifts.

Our study involved a statistical analysis of a large sample of haloes at different redshifts. This is useful for discovering and characterising correlations in the data, but it has the disadvantage of being difficult to isolate the physical processes responsible for these trends. A next step involves studying the evolution of individual haloes and identifying the non-linear processes that affect the halo spin evolution in the context of the cosmic web. Such a study is challenging due to at least two aspects: i) halo spin growth includes an intrinsic level of stochasticity due to the hierarchical and anisotropic nature of halo formation \citep[e.g. see][]{Contreras2017}, and ii) the correlation between halo spin and the web environment is rather weak and thus a large number of objects need to be studied to reliably identify the relevant processes. Nonetheless, despite these challenges, studying the formation history of individual haloes is key to understand halo spin acquisition and its relation to the cosmic web.

\section*{Acknowledgements}
We are grateful to Bernard Jones, Enn Saar, Cristiano Porciani, Noam Libeskind and Joss Bland-Hawthorn for encouraging and insightful discussions. 
PGV and ET acknowledge the support by ETAg grant PRG1006 and by EU through the ERDF CoE grant TK133.
MC acknowledges support by the EU Horizon 2020 research and innovation programme under a Marie Sk{\l}odowska-Curie grant agreement 794474 (DancingGalaxies) and by the ERC Advanced Investigator grant, DMIDAS [GA 786910].
This work used the DiRAC@Durham facility managed by the Institute for Computational Cosmology on behalf of the STFC DiRAC HPC Facility (www.dirac.ac.uk). The equipment was funded by BEIS capital funding via STFC capital grants ST/K00042X/1, ST/P002293/1, ST/R002371/1 and ST/S002502/1, Durham University and STFC operations grant ST/R000832/1. DiRAC is part of the National e-Infrastructure.

\subsection*{Data Availability}
The data underlying this article will be shared on reasonable request to the corresponding author.

\bibliographystyle{mnras}
\bibliography{bibliography}





\bsp 
\label{lastpage}

\end{document}